\documentclass[iicol, sn-nature]{sn-jnl}

\usepackage{graphicx}%
\usepackage{multirow}%
\usepackage{amsmath,amssymb,amsfonts}%
\usepackage{amsthm}%
\usepackage{mathrsfs}%
\usepackage[title]{appendix}%
\usepackage{xcolor}%
\usepackage{textcomp}%
\usepackage{manyfoot}%
\usepackage{booktabs}%
\usepackage{algorithm}%
\usepackage{algorithmicx}%
\usepackage{algpseudocode}%
\usepackage{listings}
\usepackage{siunitx}
\usepackage{dirtytalk}
\usepackage{xr}
\usepackage[sort&compress]{cleveref}
\usepackage[utf8]{inputenc}
\DeclareUnicodeCharacter{2009}{\,}

\theoremstyle{thmstyleone}%
\theoremstyle{thmstyletwo}%

\theoremstyle{thmstylethree}%

\begin{document}

\title[Article Title]{A two-dimensional 10-qubit array in germanium with robust and localised qubit control}

\author*[1]{\fnm{Valentin} \sur{John}} \email{V.John@tudelft.nl}
\equalcont{These authors contributed equally to this work.}

\author[1]{\fnm{Cécile} \spfx{X.} \sur{Yu}}
\equalcont{These authors contributed equally to this work.}

\author[1]{\fnm{Barnaby} \spfx{van} \sur{Straaten}}

\author[3]{\fnm{Esteban} \spfx{A.} \sur{Rodríguez-Mena}}

\author[3]{\fnm{Mauricio} \sur{Rodríguez}}

\author[2]{\fnm{Stefan} \sur{Oosterhout}}

\author[1]{\fnm{Lucas} \spfx{E. A.} \sur{Stehouwer}}

\author[1]{\fnm{Giordano} \sur{Scappucci}}

\author[1]{\fnm{Stefano} \sur{Bosco}}

\author[1]{\fnm{Maximilian} \sur{Rimbach-Russ}}

\author[3]{\fnm{Yann-Michel} \sur{Niquet}}

\author[1]{\fnm{Francesco} \sur{Borsoi}}

\author*[1]{\fnm{Menno} \sur{Veldhorst}} \email{M.Veldhorst@tudelft.nl}

\affil[1]{\orgdiv{QuTech and Kavli Institute of Nanoscience}, \orgname{Delft University of Technology}, \orgaddress{\street{P.O. Box 5046}, \city{Delft}, \postcode{2600 GA}, \state{Delft}, \country{The Netherlands}}}

\affil[2]{\orgname{QuTech and Netherlands Organisation for Applied Scientific Research}, \orgaddress{ \city{Delft}, \postcode{2628 CK}, \country{The Netherlands}}}

\affil[3]{\orgname{Université Grenoble Alpes, CEA, IRIG-MEM-L\_Sim}, \orgaddress{ \city{Grenoble}, \postcode{38000}, \country{France}}}


\abstract{
Quantum computers require the systematic operation of qubits with high fidelity. For holes in germanium, the spin-orbit interaction allows for \textit{in situ} electric  fast and high-fidelity qubit gates. However, the interaction also causes a large qubit variability due to strong g-tensor anisotropy and dependence on the environment. 
Here, we leverage advances in material growth, device fabrication, and qubit control to realise a two-dimensional 10-spin qubit array, with qubits coupled up to four neighbours that can be controlled with high fidelity. 
By exploring the large parameter space of gate voltages and quantum dot occupancies, we demonstrate that plunger gate driving in the three-hole occupation enhances electric-dipole spin resonance (EDSR), creating a highly localised qubit drive. Our findings, confirmed with analytical and numerical models, highlight the crucial role of intradot Coulomb interaction and magnetic field direction. Furthermore, the ability to engineer qubits for robust control is a key asset for further scaling.
}


\maketitle

\section{Introduction}\label{sec1}

Semiconductor spin qubits have seen significant progress over the last few years, with four-qubit and six-qubit quantum processors demonstrated across different platforms and encodings \cite{Hendrickx2021AProcessor, Philips2022UniversalSilicon, Zhang2024UniversalQubits, Thorvaldson2024GroversThreshold}. In a drive to scale beyond these systems, larger quantum dot arrays have been explored showcasing charge tune-up in a 4$\times$4 quantum dot array using a crossbar architecture \cite{Borsoi2024SharedArray}, qubit characterisation of a two-dimensional 10-quantum dot array by coherent single spin shuttling \cite{Wang2024OperatingSpins}, and demonstration of a linear array comprising 12 qubits \cite{George202412-spin-qubitLine}.

Hole spin qubits in planar strained Ge/SiGe heterostructures emerged as a compelling direction that can offer electrical control, fast Rabi driving, long coherence times, and absence of valley degree of freedom \cite{Scappucci2020TheRoute}. The strong anisotropy of the g-tensor for germanium hole spins creates sweet spots and lines, where qubit quality is maximized \cite{Wang2021OptimalQubits,Hendrickx2024Sweet-spotSensitivity, Carballido2024Compromise-FreeCoherence, Bassi2024OptimalQubits}. However, these optimal locations are hard to predict and sensitive to magnetic field angle variations \cite{Abadillo-Uriel2023Hole-SpinInteractions, Mauro2024DephasingSweetSpot}, which differ across quantum dots due to device-specific and cooldown-dependent potential landscapes. Optimizing qubit performance at a fixed magnetic field direction is thus crucial for consistent high-fidelity operation across many qubits.

Most experiments in germanium extensively focused on quantum dots with single-hole occupation \cite{Hendrickx2021AProcessor, Jirovec2021AGe,  vanRiggelen2022PhaseQubits, Lawrie2023SimultaneousThreshold, Wang2023ProbingSimulator, John2024BichromaticQubits, vanRiggelen-Doelman2024CoherentDots, Hendrickx2024Sweet-spotSensitivity, Wang2024OperatingSpins}. Interestingly, silicon metal-oxide-semiconductor (SiMOS) devices experience increased Rabi driving efficiencies for electron spins when occupying higher orbital states, such as the \textit{p} or \textit{d} shells, rather than the \textit{s} shell \cite{Leon2020CoherentDot}. 
A key question, therefore, is understanding the role of hole occupancy and the impact of the driving gate on the performance and crosstalk of EDSR in buried quantum wells with spin-orbit interaction.

In this work, we investigate a two-dimensional 10-quantum dot device hosting 10 qubits, with two central qubits each connected to four different neighbours in four directions. We systematically evaluate the EDSR driving efficiency of each qubit for quantum dots occupied with one, three, and five holes, to assess how charge configuration influences driving mechanisms. This analysis is extended across all 22 available gates in the device for each qubit, which gives us insights about the locality and crosstalk of EDSR driving. Additionally to probe the variation in noise sensitivity in each configuration, we perform longitudinal spin-electric susceptibility (LSES) measurements by analysing the changes in resonance frequency as a function of gate voltages under different charge configurations which is closely related to the qubit coherence. Crucially, we find that it is possible to have systematic efficient driving with limited crosstalk when operating with three-hole occupancy using the plunger gate.

\begin{figure}[htbp!]
    \centering
    \includegraphics[width=76 mm]{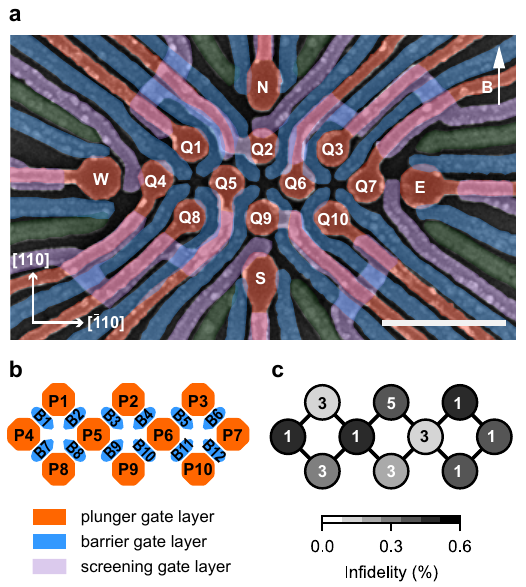}
    \caption{\textbf{A high-fidelity 10-spin-qubit array in germanium.} 
        \textbf{a,}~A false-coloured scanning electron microscope image of a nominally identical device, with the 10 quantum dot plunger gates highlighted in orange and the twelve barrier gates shown in blue. Four single hole transistors labelled as N, E, W and S are located at the edge of the array. The 10 qubits are labelled as Q1-Q10. The applied magnetic field is \SI{41.4}{mT}. The scale bar on the bottom right represents 500 nm.
        \textbf{b,}~Simplified gate layout of the quantum dot array where the plunger gates are labelled as P1-P10 and the barrier gates as B1-B12.
        \textbf{c,}~Randomised benchmarking single-qubit gate infidelities with the corresponding charge occupation of the 10 quantum dots annotated. 
    }
    \label{fig:device}
\end{figure}

\section{The two-dimensional 10-spin qubit array}

 \autoref{fig:device}a displays our device, comprising 10 quantum dots (QDs) arranged in a 3-4-3 configuration, and four charge sensors located at the cardinal points of the array, similarly to the device layout described in Refs. \cite{Wang2024OperatingSpins} and \cite{Rao2024MAViS:Arrays}. In this work, we have fabricated the quantum device on a Ge/SiGe heterostructure grown on a germanium wafer \cite{Stehouwer2023GermaniumDisorder}, exhibiting a high mobility of $3.4(1) \times 10^6$ cm$^2$/Vs, indicating a uniform and low-noise potential landscape for quantum dot arrays  \cite{2024StehouwerLowNoise}. The quantum dots are defined and operated using plunger and barrier gates, as illustrated in \autoref{fig:device}b.  A magnetic field of \SI{41.4}{\milli \tesla}, tilted approximately 2-3 degrees from the in-plane orientation \cite{2024StehouwerLowNoise}, is applied to the system.  This low magnetic field amplitude enables us to perform qubit control with arbitrary waveform generators at low frequencies up to \SI{400}{\mega \hertz} without the need of IQ-modulation with additional microwave generators.

The 10-QD array is tuned to a dense charge configuration, with an odd number of holes at each QD site, defining 10 qubits labelled Q1–Q10. Each qubit is initialised and readout pairwise using Pauli spin blockade with a nearby charge sensor \cite{Ono2002CurrentSystem, Fransson2006PauliDots}. \autoref{fig:device}c shows the occupation of each quantum dot in the initial tune-up, along with the corresponding single-qubit gate infidelity, all below 0.6\%, obtained through randomized benchmarking \cite{Knill2008RandomizedGates}. An in-depth noise analysis detailed in Ref. \cite{2024StehouwerLowNoise} indicates that the qubit performance is bounded by a hyperfine-limited $T_\mathrm{2}^*$ of approximately \SI{2}{\micro \second} arising from the out-of-plane component of the magnetic field, which could be alleviated by using purified germanium \cite{Sigillito2015ElectronGermanium}.
 We also demonstrate tunable exchange interactions between neighbouring qubit pairs (see Supplementary Note S5), realising a 10-qubit system with increased connectivity up to four nearest neighbours. 

\begin{figure*}[htbp!]
    \centering
    \includegraphics[width=\textwidth]{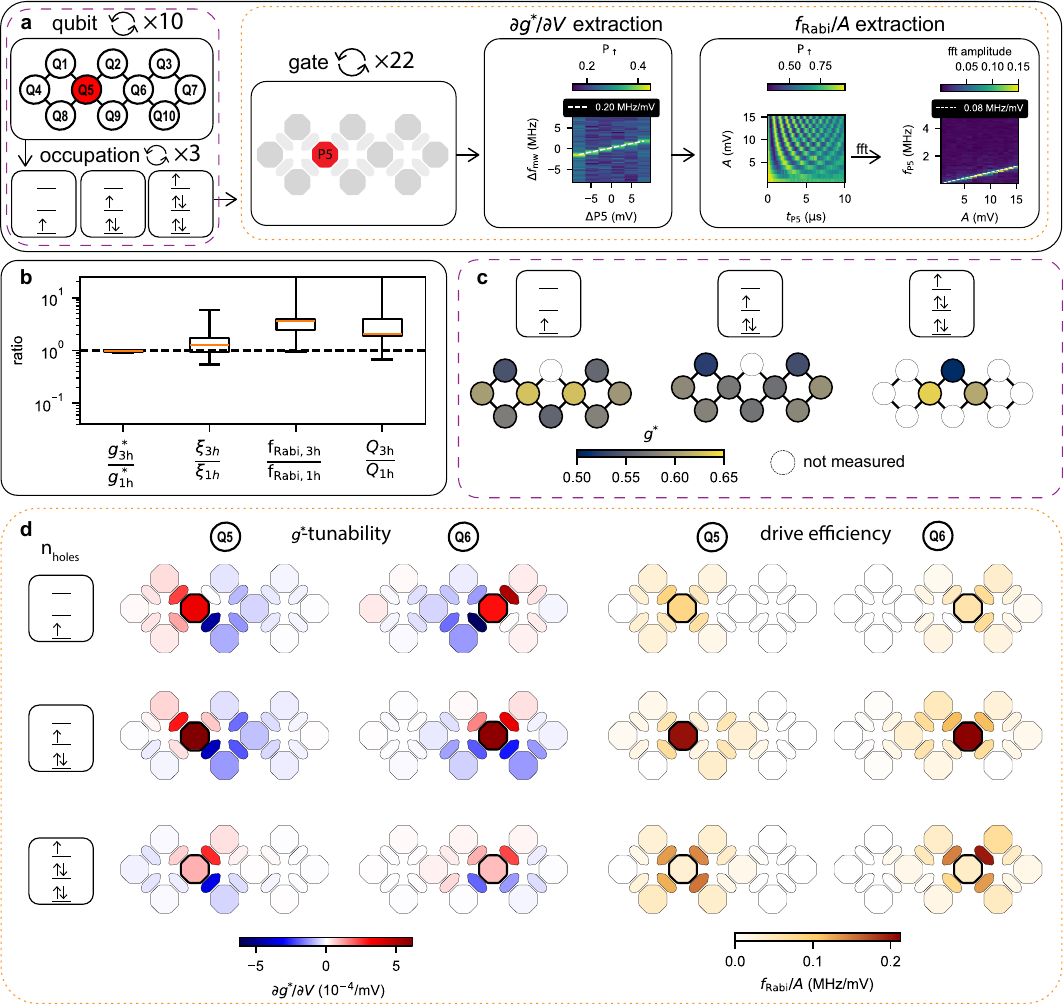}
    \caption{\textbf{Statistical analysis of the 10-spin qubit arrray.} 
        \textbf{a,}~Flow diagram of $g^*$-tunability and driving efficiency extraction. After selecting one of the qubits, Q1-Q10, and looping over the one-, three-, and five-hole occupation, the $g^*$-tunability and drive efficiencies are extracted sequentially for each of the 10 plunger and 12 barrier gates. For the $g^*$-tunability each of the gates is scanned in the range of $\pm 8 \rm{mV}$ while sweeping the microwave frequency across the corresponding qubit frequency on a dedicated fixed qubit gate. By fitting the peak in the recorded signal, the frequency slope can by determined in MHz/mV, which can be converted into a g-factor slope in 1/mV. For the driving efficiency the amplitude is swept from 1 to 15 mV, while applying a microwave pulse on each of the gates. By performing a fast-Fourier transform and fitting the dominant frequency contribution with a linear fit, the driving slope in MHz/ mV can be extracted.
        \textbf{b,}~Boxplots in Spear style containing the ratios of $g^*$, LSES, driving efficiency, and quality factor in the three- and one-hole occupation. Each boxplot contains 9 data points with all 10 qubits except Q2.
        \textbf{c,}~$g^*$ of all 10 qubits in the one-, three-, and five-hole occupation.
        \textbf{d, }~$g^*$-tunability and driving efficiency for qubit Q5 and Q6 as a function of all 10 plunger and 12 barrier gates. Each row corresponds to a different hole occupation of one, three, and five respectively. 
    }
    \label{fig:statistics}
\end{figure*}

\section{Qubit drive efficiency and tunability}

This two-dimensional 10-qubit array provides a sufficiently large and robust platform to gather a comprehensive dataset on the effects of varying qubit sites and hole occupancies while avoiding device-to-device variability. By systematically performing the measurement protocol shown in \autoref{fig:statistics}a, we characterise the LSES and driving efficiency across all qubits with one, three, and five-hole occupancy.

These driving properties are intimately linked to the sensitivity of the g-tensor to the electrostatics and its environment \cite{Crippa2017ElectricalQubits, Piot2022SweetSpot}. Indeed, Rabi oscillations are governed by modulation of the transverse component of the g-tensor through AC gate voltages, while the LSES measures the gate ability to tune the longitudinal component, which influences qubit coherence in the charge-noise-limited regime via  $T_2^* \propto 1/\xi$, with  $\xi = \sqrt{\sum_{\mathrm{gate}}(\partial{g^*} / \partial{V_{\mathrm{gate}}})^2}$ \cite{Piot2022SweetSpot, Hendrickx2024Sweet-spotSensitivity}. 
Here, the effective g-factor for a given magnetic field is expressed as $g^*=|\mathbf{g}\vec{b}|$, with $\mathbf{g}$ representing the g-tensor and $\vec{b}=\vec{B}/|\vec{B}|$ the normalised magnetic field direction. $\xi$ is the total g-factor susceptibility over all the gates of the device, assuming uncorrelated g-factor tunability between different gates. The interplay between driving efficiency and longitudinal susceptibility can be captured by a quality factor, defined as $Q = f_\mathrm{Rabi}/(A_\mathrm{Rabi}\cdot \xi)$, enabling identification of operational sweet spots and their dependence on hole configurations.

\autoref{fig:statistics}b summarises qubit statistics collected across the 10 qubits for single and triple-hole occupations, visualizing their ratios in a boxplot. As the hole occupancy increases from one to three, both $g^*$ and $\xi$ show minimal variation, while the plunger driving efficiency improves  by a median factor of 3.6. With a modest median increase of 1.3 in $\xi$, the quality factor improves by a median factor of 2.5\footnote{Here, we refer to the median values, as the average is skewed by Q3, which is barely driven by the plunger gate in the single-hole occupation regime.}.
Notably, the whisker representing the Rabi frequency ratio for single- and triple-hole occupancies extends towards infinity, as no measurable driving of Q3 using the gate P3 was observed in the single-hole occupancy within the applied voltage amplitude range, which confirms the importance of investigating the gate dependence of EDSR in large qubit arrays. 

The underlying data of the g-factor variability is visualised in \autoref{fig:statistics}c across different hole occupancies and qubit sites. Considering the large g-tensor anisotropy, these data reveal fairly minimal relative variation in g-factors across 10 qubits within a single device and different hole configurations, with an average g-factor of $0.58 \pm 0.03$. The small variability of the g-factor can be attributed to the slightly out-of-plane magnetic field, since the out-of-plane component of the g-tensor is much larger and varies much less relative to the in-plane principal g-factors.

Exemplary data for the central qubits, Q5 and Q6, which are measured across all three hole occupancies, are shown in \autoref{fig:statistics}d. In the following, we refer to plunger drive when the qubit is driven with a plunger gate, and barrier drive when a barrier gate is used. The data show a distinct increase in qubit plunger drive as the hole occupation increases from one to three, while the contributions from the barriers remain approximately unchanged. However, increasing the hole occupation to five reverses this pattern, with the qubit plunger drive becoming significantly weaker while the barrier drives become stronger. The observed $g^*$-tunability patterns include barriers exhibiting both negative and positive $\partial{g^*} / \partial{V_{\textrm{gate}}}$, but the relative positions of these barriers do not reveal clear trends across the full array, making it challenging to identify their origin. Two distinct patterns are notable though. Barriers along the diagonal often have approximately opposite values, while barriers at the top or bottom of the array often share the same sign. The $\partial{g^*} / \partial{V_{\textrm{gate}}}$ associated with the qubit plunger is always positive, but its magnitude is comparable to that of the associated barriers. These patterns are supported and explained by numerical simulations in Supplementary Note S12. 

The trend of increased top plunger driving efficiency from one to three holes is observed in eight of the nine measured qubits (Q2 has only been measured in the five-hole regime). The exception is qubit Q4, which exhibits a constant plunger drive efficiency from one- to three-hole occupation. This behaviour may be related to the small charging energy of the dot, resulting in a larger wavefunction that already enables efficient drive in the single-hole regime. The complete dataset for $g^*$-tunability and driving efficiency across all qubits and gates is provided in Supplementary Note S8 and S9. Overall, the three-hole regime is a more favourable regime for operation, as the driving mechanisms are more robust, with importantly no instances of zero driving, unlike in the one-hole regime.

\begin{figure*}[btp!]
    \centering
    \includegraphics[width=\textwidth]{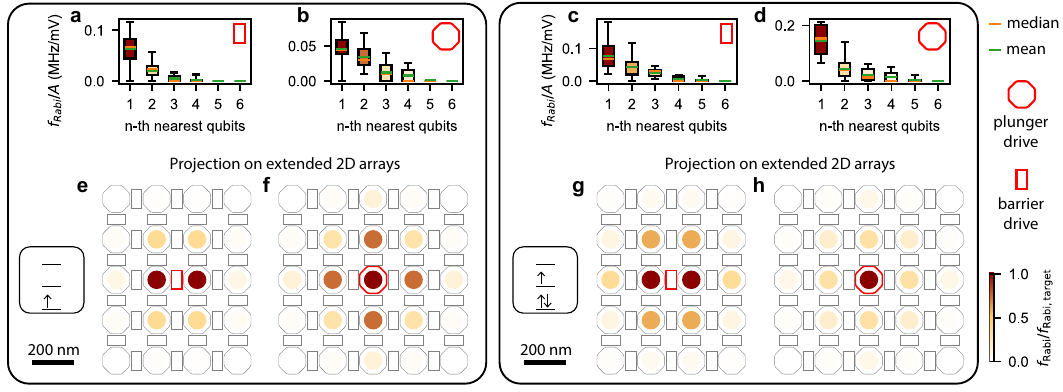}
    \caption{\textbf{Rabi driving locality}
    \textbf{a–d,} Boxplots depicting the Rabi driving efficiency for all gates up to the sixth nearest qubit with indicated mean (green line) and median value (orange line). Data are shown for (\textbf{a}) barrier drive in the single-hole occupation, (\textbf{b}) plunger drive in the single-hole occupation, (\textbf{c}) barrier drive in the three-hole occupation, and (\textbf{d}) plunger drive in the three-hole occupation. The colour of each boxplot represents the Rabi frequency normalized to that of the target qubit.
    \textbf{e–h,} Projection of the normalized Rabi frequency onto an extended, densely populated 2D spin qubit array. Data are shown for  barrier drive in the single-hole occupation (\textbf{e}), plunger drive in the single-hole occupation (\textbf{f}), barrier drive in the three-hole occupation (\textbf{g}), and plunger drive in the three-hole occupation (\textbf{h}).
    }
    \label{fig:locality}
\end{figure*}

\section{Driving locality in extended qubit arrays}

EDSR driving is commonly expected to be a local phenomenon due to the localized nature of the electric field \cite{Golovach2006Electric-dipole-inducedDots, Nowack2007CoherentFields}. Here, we leverage the densely occupied 10-qubit system to quantify this assertion. To do so, we analyse the acquired data by focusing on the driving efficiency of each gate to each qubit. Specifically, we evaluate the driving locality when applying a microwave pulse to any of the 22 available plunger and barrier gates. In this analysis, we consider four distinct cases involving either barrier or plunger drive and either one- or three-hole occupation. The five-hole occupation is excluded in this study due to insufficient data across the array. For each driving gate, the corresponding target qubit is defined as the qubit closest to the driving gate.

In \autoref{fig:locality}, driving efficiency is categorised by the physical distance of each gate to each qubit. Independent of direction, we then define nearest neighbours based on this physical proximity. Driving efficiency is quantified by the averaged results for all \textit{n}-th nearest qubits over all driving gates. \autoref{fig:locality}a–d present the Rabi driving efficiencies as boxplots up to the sixth nearest neighbour for both barrier and plunger drive, with a corresponding maximum physical distance of 550 nm in the device plane (all distances of the \textit{n}-th nearest qubits and their ranks are listed in Supplementary Table S4 and S5). 
To evaluate drive locality, we express our data in terms of the normalized driving efficiency $f_{\text{Rabi}}/f_{\text{Rabi,target}}$, obtained by normalizing the Rabi driving efficiency relative to that of the target qubit.
Lower normalized driving efficiencies for distant qubits indicate less cross-talk and more localized driving. For qubits located beyond the sixth nearest neighbour, driving efficiency falls below \SI{0.01}{MHz/mV}, which is below the sensitivity of the measurement within the range of applied drive amplitudes. We generally observe an expected decrease in driving efficiency for larger distances in both one- and three-hole cases, for both plunger and barrier drives. The drop in mean driving efficiency from the first to the second nearest qubit is largest for the three-hole plunger drive and single-hole barrier drive.

To illustrate the impact of cross-talk, we project the measured results onto an extended densely populated 2D spin qubit array. When driving a target qubit with a specific gate, the \textit{n}-th nearest qubits are colour-coded according to the mean of the normalized Rabi efficiency measured experimentally and presented in the boxplots of \autoref{fig:locality}a–d. This visualization highlights how much each qubit would be affected when driven using a plunger or barrier gate under single- and triple-hole occupations.

\autoref{fig:locality}e and g show the projected cross-talk for barrier driving. By design, barrier gates drive two nearest-neighbour qubits equally, resulting in pronounced cross-talk between them. While cross-talk is slightly stronger in the single-hole occupation, driving efficiencies are slightly lower. Barrier drive can still be advantageous compared to single-hole plunger drive, as the latter may yield negligible or zero driving efficiencies, necessitating large drive amplitudes and increasing normalized efficiencies for next-nearest neighbours. As shown in \autoref{fig:locality}f, the single-hole plunger drive induces noticeable effects on next-nearest qubits.

The three-hole plunger drive improves driving efficiency while reducing normalized driving efficiencies for next-nearest qubits. Unlike in the single-hole occupation, non-driving regimes are avoided entirely in the three-hole case. This configuration achieves minimal cross-talk while maintaining the highest driving efficiencies, making it here the most favourable driving scheme in terms of efficiency and cross-talk mitigation in a 2D array with dense occupation.

\section{Modelling of single- and multi-hole quantum dots}

The improvement of driving efficiency in the three-hole regime can be well captured by a phenomenological model, including only the two lowest available orbital levels, which are spin doublets with spin $S_z=\pm 1/2$. In the weak interaction case, these states are gapped by an orbital energy $\mathcal{E}$, and are coupled by Coulomb interactions that provide an inter-orbital exchange tunnelling energy $|t|\ll |\mathcal{E}|$.
As the orbitals have different wavefunctions, these states also have different g-factors $g_{1}$ and $g_{2}$. We define the difference in g-factors as $\delta g=(g_{1}-g_{2})/2$.

This model is analogous to that of flopping-mode qubits~\cite{Benito2019Electric-fieldQubit,Croot2020Flopping-modeResonance} under the exchange of orbital and positional degrees, where the presence of an orbital state nearby significantly enhances driving efficiency.
We find that the g-tensor of the ground state 
 $g_{e}\approx g_1- { t^2}  \delta g/{2\mathcal{E}^2}$ has two contributions: a single-particle term  $g_1$ and an interaction correction  $\propto t^2/\mathcal{E}^2$. 
The resonant Rabi frequency depends on the change of the $g$-tensor caused by the driving gate potential 
\begin{align}
        f_R = \left|\Vec{f}_R^{SP} + \Vec{f}_R^{MB}\right|,
\end{align}
and can be decomposed into the single-particle (SP) magnetic  response $|\Vec{f}_R^{SP}|\propto  \partial_{V_k}g_1$ caused by an electric modulation of the single-particle g-tensor and the many-body (MB) interaction effect $ |\Vec{f}_R^{MB}|\propto  \partial_{V_k}t^2/\mathcal{E}^2$ caused by the modulation of the hybridization.
The prefactor $t^2/\mathcal{E}^2\propto \ell^2$ is highly sensitive to changes in the confinement potential that affects the dot size $\ell$~\cite{DiVincenzo1999CoupledGates}. As a result, when the driving gate is the top plunger gate, $f_R^{MB}$ dominates the response, therefore driving efficiency is improved. 
On the other hand, $t^2/\mathcal{E}^2$  is only weakly sensitive to electric fields that shifts the wavefunction, and thus when the driving gate is a side barrier gate, the many-body correction is negligible and the driving is only determined by $f_R^{SP}$, which is typically small for roughly circular dots, and is only enhanced for squeezed dot shapes~\cite{Bosco2021SqueezedPower}. More details of this model can be found in Supplementary Note S12.

In addition to the phenomenological model, we also perform full configuration interaction simulations based on the four-band Luttinger-Kohn model in both single- and triple-hole quantum dots that are in good agreement with our experimental results. In these simulations, we explore different dot parametrisations to reproduce the typical patterns arising from the LSES and EDSR data of each qubit to each gate (see Supplementary Note S12). This study shows the dependence of the drive efficiency on the magnetic field angle in both single- and triple-hole regimes. Particularly, the many-body term $f_{res}^{MB}$ becomes negligible when the magnetic field is well-aligned in-plane (see Supplementary Note S12). This suggests that for fully in-plane magnetic fields, transitioning from single- to triple-hole occupation does not improve plunger drive efficiency.

\section{Conclusions}
In this work, we show that spin qubits can be configured in a two-dimensional array, with qubits connected up to four neighbouring qubits, and operated with high single-qubit gate fidelity. We have explored the driving of all 10 qubits, to understand the locality of EDSR and obtain best and robust driving conditions. Our findings demonstrate that the pronounced g-tensor anisotropy in germanium can be exploited to engineer a set of qubit properties by means of hole occupation in conjunction with the magnetic field direction. In particular, we demonstrate that a slight out-of-plane magnetic field can enable uniform g-factors and in the three-hole occupation result in reproducible and dominant plunger-driven Rabi frequencies. This regime shows particular promise for achieving highly localised and systematic qubit control, critical for scaling to multi-qubit systems. Our work also demonstrates the ability to engineer qubit properties in semiconductor qubit arrays and highlights the need to assess the qubit performance as function of magnetic field strength and angle, quantum dot shape, and hole occupancy, to tailor these parameters to the specific architecture.

\section*{Acknowledgements}
We acknowledge Floor van Riggelen-Doelman for support in the data analysis, Chien-An Wang for fruitful discussions and Sander de Snoo for software support. 
This research was supported by the European Union through the ERC Starting Grant QUIST (850641), the Horizon 2020 research and innovation programme under the Grant Agreement No. 951852 (QLSI) and the Horizon Europe Framework Programme under grant agreement No. 101069515 (IGNITE). This research was sponsored in part by the Army Research Office (ARO) under Award No. W911NF-23-1-0110. The views, conclusions, and recommendations contained in this document are those of the authors and are not necessarily endorsed nor should they be interpreted as representing the official policies, either expressed or implied, of the Army Research Office (ARO) or the U.S. Government. The U.S. Government is authorized to reproduce and distribute reprints for Government purposes notwithstanding any copyright notation herein. F.B. acknowledges support from the NWO through the National Growth Fund program Quantum Delta NL (grant NGF.1582.22.001). M.R-R. acknowledges support from NWO under Veni Grant (VI.Veni.212.223)

\section*{Author contributions}
V.J., C.X.Y., F.B. and B.v.S. conducted the experiments and V.J. performed the analysis. S.O. fabricated the device, L.E.A.S. and G.S. supplied the heterostructures. E.R-M., M.R., S.B., M.R-R. and Y-M.N. performed the simulation and theoretical analysis, and S.B., M.R.-R. and Y-M.N. supervised the theory section. V.J., C.X.Y. and M.V. wrote the manuscript with input of all authors. M.V. supervised the project.


\section*{Declarations}
M.V. and G.S. are founding advisors of Groove Quantum BV and declare equity interests. The remaining authors declare that they have no competing interests.

\section*{Data availability}

The data supporting the findings of this study are openly available in the 4TU.ResearchData repository under the DOI: 10.4121/5ee5b0d3-e838-478e-990d-02c50b75eeab.

\bibliography{mendeley_references}

\end{document}



\newcommand{\colorcell}[1]{%
    \ifnum#1<3 \cellcolor{blue!20}#1\else
    \ifnum#1<7 \cellcolor{yellow!20}#1\else
    \cellcolor{red!20}#1\fi\fi}

\newcommand{\colorcelldis}[1]{%
    \FPeval\result{round(#1, 2)} 
    \ifdim\result pt<0.25pt \cellcolor{blue!20}\result\else
    \ifdim\result pt<0.56pt \cellcolor{yellow!20}\result\else
    \cellcolor{red!20}\result\fi\fi}

\renewcommand{\thepage}{S\arabic{page}} 
\renewcommand{\thesection}{Suppl. Note \arabic{section}}  

\setcounter{figure}{0}
\setcounter{page}{1}
\setcounter{section}{0}
\setcounter{table}{0}

\renewcommand{\figurename}{\textbf{Supplementary Figure}}
\renewcommand{\thefigure}{\textbf{S\arabic{figure}}}

\renewcommand{\tablename}{\textbf{Supplementary Table}}
\renewcommand{\thetable}{\textbf{S\arabic{table}}}

\renewcommand{\bibnumfmt}[1]{[S#1]}
\renewcommand{\citenumfont}[1]{S#1}
\onecolumngrid

\tableofcontents
\newpage
\section{Materials and Methods}

The device is fabricated on a Ge/SiGe heterostructure with a 16 nm germanium quantum well buried 55 nm below the semiconductor/oxide interface on a germanium substrate as described in Ref.\cite{2024StehouwerLowNoise}. The ohmic contacts are created first through patterning and platinium depostion. For the gate stack,the barrier gate layer (20nm thick) is deposited, followed by the screening gate layer (30nm thick), and finally the plunger layer (40nm thick). All the gate layers are made of palladium, deposited at room temperature. The barrier gates are separated from the heterostructure by a 7 nm thick aluminium oxide and a 5nm oxide is deposited after the barrier and the screening layers. 

The experiments are performed in a LD-400 Bluefors dilution fridge equipped with a solenoid superconducting magnet. Details of the set-up can be found in Ref.\cite{Wang2024OperatingSpins}.

\begin{figure*}[htb!]
	\centering
	\includegraphics{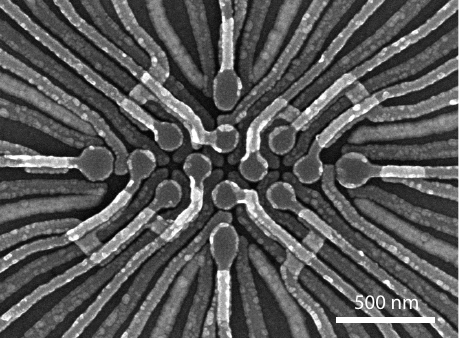}
	\caption{Scanning electron microscope image of a device nominally identical to the one utilised in the experiments without any false colouring.}
	\label{fig:sem_no_colour}
\end{figure*}

\section{Gate virtualisation}

We use a set of virtual gates, defined in software, as outlined in \autoref{tab:virtual_steps}. The first layer of virtualisation addresses crosstalk between gates and sensors as well as interactions between nearby quantum dots. This ensures that adjustments to the plunger or barrier gates do not shift the position of the charge sensor’s Coulomb peak or change nearby dots potentials. In the third layer of virtualisation, the charging voltage of each dot is normalised by rescaling the strength of its corresponding plunger. Finally, the third layer virtualizes the barrier gates, enabling independent tuning of the exchange interaction between dots without changing their charge states. For a complete gate virtualisation method, readers may refer to Ref.\cite{Rao2024MAViS:Arrays}.

\begin{table*}[h]
\renewcommand{\arraystretch}{1.1}
\renewcommand{\tabcolsep}{2pt}
\caption{\label{tab:virtual_steps}
Table illustrating the layers of our virtualization approach. \textbf{S} denotes sensor plunger gates, \textbf{P} (\textbf{B}) defines quantum dot plunger (barrier) gates.}
\begin{ruledtabular}
\begin{tabular}{>{\centering\arraybackslash}m{0.8cm}p{9.0cm}c>{\centering\arraybackslash}m{1.0cm}}
\multirow{2}{0.8cm}{\centering Virt. layer} & 
\multirow{2}{9.0cm}{Description} & 
\multirow{2}{*}{Notation} \\ \\ 
\hline
1 & Charge sensor and QD compensations &  $[\mathbf{vS}, \mathbf{vP}, \mathbf{vB}]=\mathbf{M_1} \cdot [\mathbf{S}, \mathbf{P}, \mathbf{B}]$ \\ 
2 & Normalisation of plungers with uniform charging voltages & $\mathbf{N}=\mathbf{M_2} \cdot \mathbf{vP}$  \\ 
3 & Barriers to QDs  & $\mathbf{J}=\mathbf{M_3} \cdot [\mathbf{vB}, \mathbf{N}]$ \\ 
\end{tabular}
\end{ruledtabular}
\end{table*}

\newpage

\section{Qubit properties in the initial hole configuration}

\begin{table}[h!]
\caption{Qubit properties in the original charge configuration at 41.4mT.}
\label{tab:horseshoe_overview}
\begin{tabular}{|c|c|c|c|c|c|c|c|c|c|c|}
\hline
\textbf{label}                             & \textbf{Q1} & \textbf{Q2} & \textbf{Q3} & \textbf{Q4} & \textbf{Q5} & \textbf{Q6} & \textbf{Q7} & \textbf{Q8} & \textbf{Q9} & \textbf{Q10} \\ \hline
\textbf{Charge occupation}                 & 3           & 5           & 1           & 1           & 1           & 3           & 1           & 3           & 3           & 1            \\ \hline
\textbf{Larmor frequency (MHz)}            & 302         & 293         & 319         & 347         & 361         & 321         & 355         & 335         & 327         & 331          \\ \hline
\textbf{g-factor}                          & 0.52        & 0.51        & 0.55        & 0.6         & 0.62        & 0.55        & 0.61        & 0.58        & 0.56        & 0.57         \\ \hline
\textbf{coherence time (ns)}               & 2381        & 2252        & 1710        & 2238        & 2041        & 2134        & 2319        & 2207        & 2018        & 1975         \\ \hline
\textbf{Maximum driving strength (MHz/mV)} & 0.15        & 0.19        & 0.05        & 0.06        & 0.08        & 0.21        & 0.13        & 0.15        & 0.09        & 0.11         \\ \hline
\textbf{Single-qubit fidelity (\%)}                     & 99.9        & 99.6        & 99.5        & 99.6        & 99.6        & 99.8        & 99.6        & 99.7        & 99.5        & 99.6         \\ \hline
\end{tabular}
\end{table}

\section{Single-qubit gate randomised benchmark}

We perform randomised benchmarking to measure the single-qubit gate fidelity of the 10 qubits in the initial hole configuration, identified in \autoref{tab:horseshoe_overview}, using the set of Clifford gates defined in \autoref{tab:clifford_set}. Details about this Clifford group can be found in Ref. \cite{Xue2019BenchmarkingDevice}.

\begin{table}[h]
\caption{\label{tab:clifford_set}
Single-qubit Clifford sequence and their composition. X$_{\pi/2}$ and Z$_{\pi/2}$ are referring to $\pi/2$ rotation around the x-axis and the z-axis, respectively, of the Bloch sphere of a single-qubit. The average number of elementary gates per Clifford composition is 2 as the Z rotation is generated by a change of qubit's reference frame in software, which makes it error-free.
}
\begin{tabular}{ m{0.1 \textwidth} | m{0.24\textwidth} }
\toprule[1pt]\midrule[0.3pt]
Clifford & Composition  \\
\hline
C$_{1}$  & X$_{\pi/2}$X$_{-\pi/2}$  \\
C$_{2}$  & X$_{\pi/2}$X$_{\pi/2}$  \\
C$_{3}$  & Z$_{-\pi/2}$X$_{\pi/2}$X$_{\pi/2}$Z$_{\pi/2}$  \\
C$_{4}$  & X$_{\pi/2}$Z$_{\pi/2}$Z$_{\pi/2}$X$_{\pi/2}$  \\
C$_{5}$  & X$_{\pi/2}$Z$_{-\pi/2}$X$_{\pi/2}$Z$_{\pi/2}$  \\
C$_{6}$  & X$_{\pi/2}$Z$_{\pi/2}$X$_{\pi/2}$Z$_{\pi/2}$  \\
C$_{7}$  & X$_{-\pi/2}$Z$_{-\pi/2}$X$_{\pi/2}$Z$_{\pi/2}$  \\
C$_{8}$  & X$_{-\pi/2}$Z$_{\pi/2}$X$_{\pi/2}$Z$_{-\pi/2}$  \\
C$_{9}$  & Z$_{-\pi/2}$X$_{\pi/2}$Z$_{\pi/2}$X$_{\pi/2}$  \\
C$_{10}$  & Z$_{-\pi/2}$X$_{\pi/2}$Z$_{\pi/2}$X$_{-\pi/2}$  \\
C$_{11}$  & Z$_{\pi/2}$X$_{\pi/2}$Z$_{-\pi/2}$X$_{\pi/2}$  \\
C$_{12}$  & Z$_{\pi/2}$X$_{\pi/2}$Z$_{-\pi/2}$X$_{-\pi/2}$  \\
C$_{13}$  & Z$_{-\pi/2}$X$_{\pi/2}$Z$_{\pi/2}$X$_{\pi/2}$Z$_{-\pi/2}$ \\
C$_{14}$  & Z$_{\pi/2}$X$_{-\pi/2}$Z$_{-\pi/2}$X$_{-\pi/2}$Z$_{\pi/2}$ \\
C$_{15}$  & X$_{\pi/2}$Z$_{\pi/2}$X$_{-\pi/2}$\\
C$_{16}$  & X$_{\pi/2}$Z$_{-\pi/2}$X$_{-\pi/2}$\\
C$_{17}$  & X$_{-\pi/2}$Z$_{\pi/2}$Z$_{\pi/2}$X$_{-\pi/2}$Z$_{-\pi/2}$ \\
C$_{18}$  & X$_{-\pi/2}$Z$_{-\pi/2}$Z$_{-\pi/2}$X$_{-\pi/2}$Z$_{\pi/2}$ \\
C$_{19}$  & X$_{\pi/2}$Z$_{-\pi/2}$X$_{\pi/2}$\\
C$_{20}$  & X$_{\pi/2}$Z$_{\pi/2}$X$_{\pi/2}$\\
C$_{21}$  & Z$_{-\pi/2}$X$_{\pi/2}$Z$_{\pi/2}$X$_{\pi/2}$Z$_{\pi/2}$\\
C$_{22}$  & Z$_{-\pi/2}$X$_{\pi/2}$Z$_{\pi/2}$X$_{-\pi/2}$Z$_{-\pi/2}$\\
C$_{23}$  & X$_{\pi/2}$X$_{\pi/2}$Z$_{\pi/2}$\\
C$_{24}$  & X$_{-\pi/2}$X$_{-\pi/2}$Z$_{-\pi/2}$\\
\midrule[0.3pt]\bottomrule[1pt]
\end{tabular}
\end{table}

\begin{figure*}[htb!]
	\centering
	\includegraphics{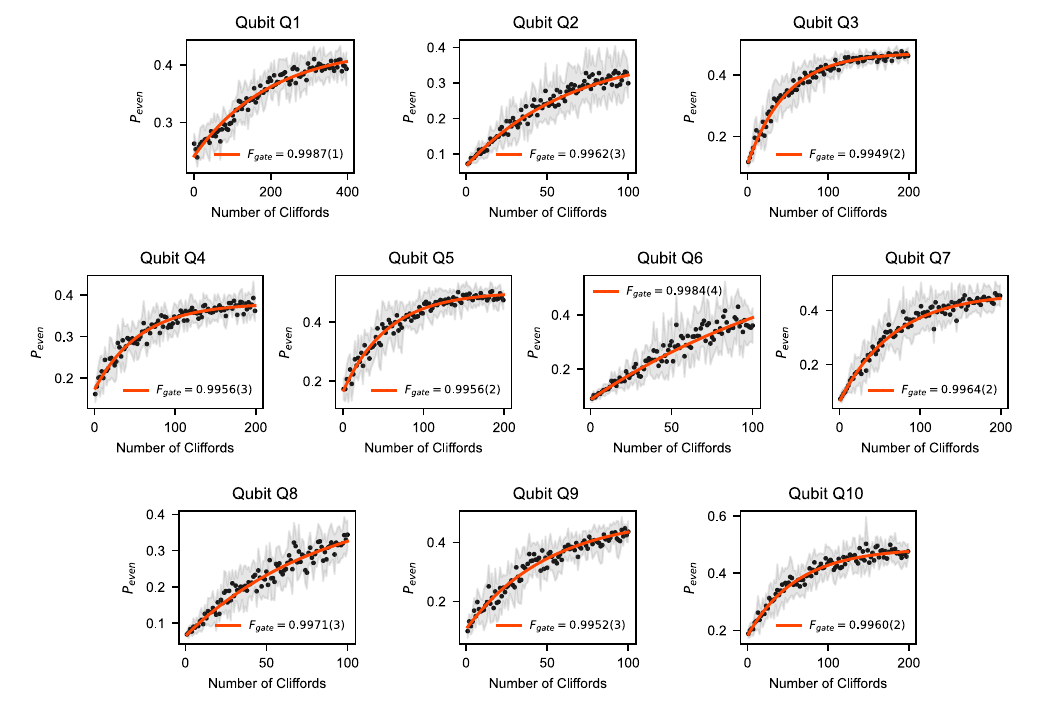}
	\caption{Single-qubit gate benchmark on the 10 qubits. The black dots correspond to the averaged randomised benchmarking data over 10 randomisations, the red line is the exponential fit to extract the gate fidelity $F_\mathrm{gate}$ and the grey area covers the standard deviation of the data. The error bar only denotes the precision of the fit. We also remark that while the sequence lengths of 100 Clifford yields saturation for fidelities below 99.4\%, larger length sequences may be needed to probe the precise fidelities of the better performing qubits.}
	\label{fig:1Q_rb}
\end{figure*}

The randomised benchmarking data for all 10 qubits are shown in \autoref{fig:1Q_rb}. We assume an exponential decay of the form $P_\mathrm{even} = aF^n+b$ where $F$ is the circuit level fidelity, $n$ is the number of Clifford operations, \textit{a} and \textit{b} are fitting parameters depending on the state preparation and measurement. The Clifford fidelity is then given by 
\begin{equation}
    F_\mathrm{C} = 1 - (1-F)/2,
\end{equation}
and the native gate fidelity is  
\begin{equation}
    F_\mathrm{gate} = 1 - (1 - F_\mathrm{C})/(2 N_{\text{avg}}),
\end{equation}
where $N_{\text{avg}}$ is the average gate number per Clifford, which for our chosen Clifford set is 2. 
\cite{Xue2019BenchmarkingDevice}.

\clearpage
\section{Exchange interaction in the 10-qubit array}

\begin{figure*}[htb!]
	\centering
	\includegraphics{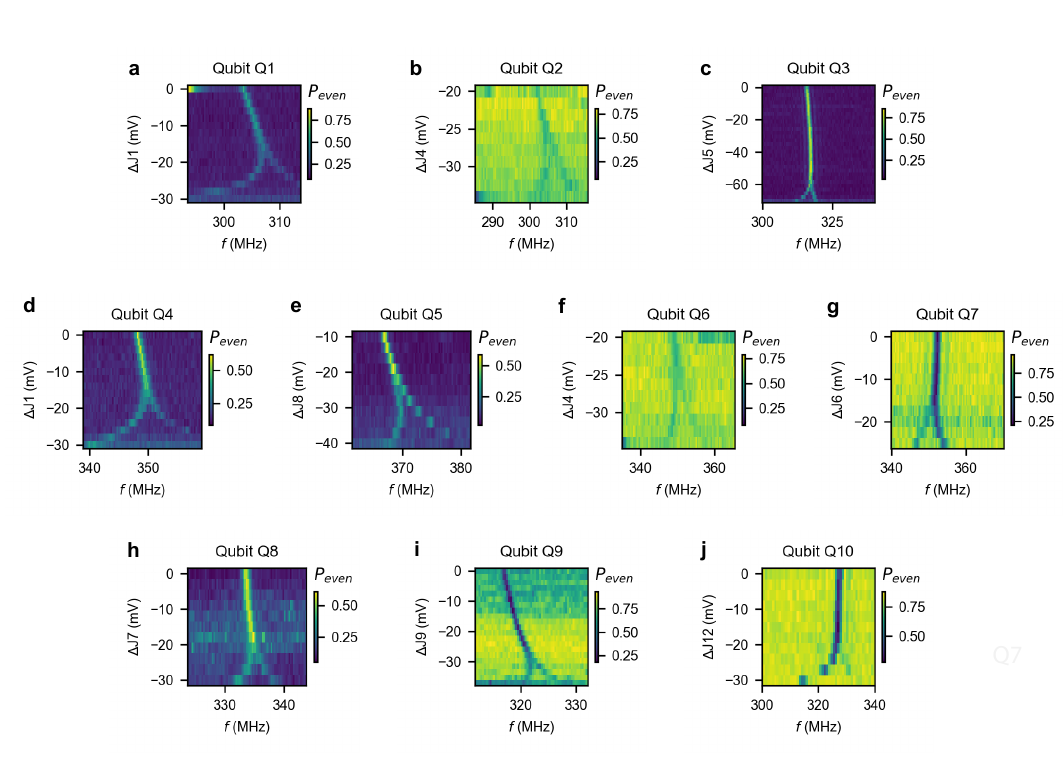}
	\caption{\textbf{a-j. Exchange splitting for all ten qubits.} The observed splitting of the qubit resonance frequency as a function of virtual barrier voltage is directly proportional to the exchange coupling between qubits.}
	\label{fig:exchange_qubit}
\end{figure*}

\begin{figure*}[htb!]
	\centering
    \includegraphics{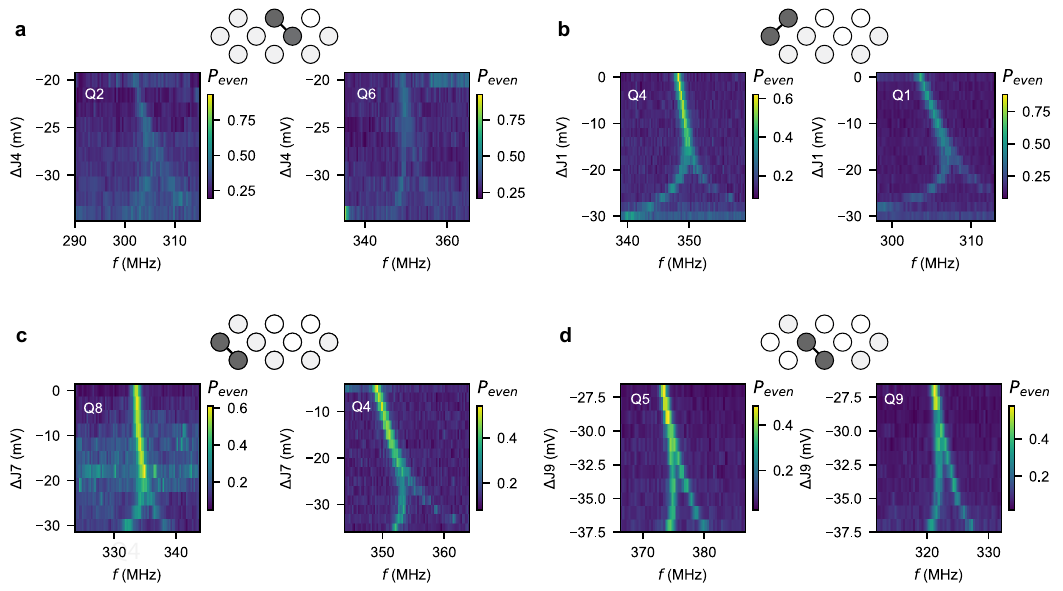}
	\caption{a-d. Examplary data of exchange interaction between qubit pairs.
	}
	\label{fig:exchange_pair}
\end{figure*}

Here we show the exchange interaction spectroscopy measurements for all the ten qubits where we observe the exchange splitting between each qubit and one neighbouring qubit, see \autoref{fig:exchange_qubit}. 
\autoref{fig:exchange_pair} shows qubit pairs, where we measured exchange interaction splitting for both qubits demonstrating the connectivity of the qubit array. 

\clearpage
\section{Physical distance from gates to qubits}

To categorise the driving efficiency of all gates to all qubits, we calculate the physical distance in the device plane from the expected qubit position at the centre of its top plunger, to any of the 22 gates. Afterwards, we rank each physical distance for plunger and barrier gates independent from each other, and assign it an integer number n corresponding to the \textit{n}-th nearest qubit-to-gate distance. We use these assigned ranks to calculate an average value for the \textit{n}-th nearest qubit for a plunger and barrier drive respectively.

\begin{table}[h!]
\caption{Physical distance from each gate to each qubit in $\mathrm{\mu m}$}
\label{tab:qubit_gate_distance}
\begin{tabularx}{\textwidth}{|c|X|X|X|X|X|X|X|X|X|X|X|X|X|X|X|X|X|X|X|X|X|X|}
\hline
             & \textbf{P1} & \textbf{P2} & \textbf{P3} & \textbf{P4} & \textbf{P5} & \textbf{P6} & \textbf{P7} & \textbf{P8} & \textbf{P9} & \textbf{P10} & \textbf{B1} & \textbf{B2} & \textbf{B3} & \textbf{B4} & \textbf{B5} & \textbf{B6} & \textbf{B7} & \textbf{B8} & \textbf{B9} & \textbf{B10} & \textbf{B11} & \textbf{B12} \\ \hline
\textbf{Q1} & \colorcelldis{0.0} & \colorcelldis{0.28} & \colorcelldis{0.55} & \colorcelldis{0.2} & \colorcelldis{0.2} & \colorcelldis{0.44} & \colorcelldis{0.70} & \colorcelldis{0.28} & \colorcelldis{0.39} & \colorcelldis{0.62} & \colorcelldis{0.1} & \colorcelldis{0.1} & \colorcelldis{0.22} & \colorcelldis{0.35} & \colorcelldis{0.49} & \colorcelldis{0.62} & \colorcelldis{0.22} & \colorcelldis{0.22} & \colorcelldis{0.29} & \colorcelldis{0.4} & \colorcelldis{0.53} & \colorcelldis{0.65} \\
\hline
\textbf{Q2} & \colorcelldis{0.28} & \colorcelldis{0.0} & \colorcelldis{0.28} & \colorcelldis{0.44} & \colorcelldis{0.2} & \colorcelldis{0.2} & \colorcelldis{0.44} & \colorcelldis{0.39} & \colorcelldis{0.28} & \colorcelldis{0.39} & \colorcelldis{0.35} & \colorcelldis{0.22} & \colorcelldis{0.1} & \colorcelldis{0.1} & \colorcelldis{0.22} & \colorcelldis{0.35} & \colorcelldis{0.4} & \colorcelldis{0.29} & \colorcelldis{0.22} & \colorcelldis{0.22} & \colorcelldis{0.29} & \colorcelldis{0.4} \\
\hline
\textbf{Q3} & \colorcelldis{0.55} & \colorcelldis{0.28} & \colorcelldis{0.0} & \colorcelldis{0.7} & \colorcelldis{0.44} & \colorcelldis{0.2} & \colorcelldis{0.2} & \colorcelldis{0.62} & \colorcelldis{0.39} & \colorcelldis{0.28} & \colorcelldis{0.62} & \colorcelldis{0.49} & \colorcelldis{0.35} & \colorcelldis{0.22} & \colorcelldis{0.1} & \colorcelldis{0.1} & \colorcelldis{0.65} & \colorcelldis{0.53} & \colorcelldis{0.4} & \colorcelldis{0.29} & \colorcelldis{0.22} & \colorcelldis{0.22} \\
\hline
\textbf{Q4} & \colorcelldis{0.2} & \colorcelldis{0.44} & \colorcelldis{0.7} & \colorcelldis{0.0} & \colorcelldis{0.28} & \colorcelldis{0.55} & \colorcelldis{0.83} & \colorcelldis{0.2} & \colorcelldis{0.44} & \colorcelldis{0.7} & \colorcelldis{0.1} & \colorcelldis{0.22} & \colorcelldis{0.35} & \colorcelldis{0.49} & \colorcelldis{0.62} & \colorcelldis{0.76} & \colorcelldis{0.1} & \colorcelldis{0.22} & \colorcelldis{0.35} & \colorcelldis{0.49} & \colorcelldis{0.62} & \colorcelldis{0.76} \\
\hline
\textbf{Q5} & \colorcelldis{0.2} & \colorcelldis{0.2} & \colorcelldis{0.44} & \colorcelldis{0.28} & \colorcelldis{0.0} & \colorcelldis{0.28} & \colorcelldis{0.55} & \colorcelldis{0.2} & \colorcelldis{0.2} & \colorcelldis{0.44} & \colorcelldis{0.22} & \colorcelldis{0.1} & \colorcelldis{0.1} & \colorcelldis{0.22} & \colorcelldis{0.35} & \colorcelldis{0.49} & \colorcelldis{0.22} & \colorcelldis{0.1} & \colorcelldis{0.1} & \colorcelldis{0.22} & \colorcelldis{0.35} & \colorcelldis{0.49} \\
\hline
\textbf{Q6} & \colorcelldis{0.44} & \colorcelldis{0.2} & \colorcelldis{0.2} & \colorcelldis{0.55} & \colorcelldis{0.28} & \colorcelldis{0.0} & \colorcelldis{0.28} & \colorcelldis{0.44} & \colorcelldis{0.2} & \colorcelldis{0.2} & \colorcelldis{0.49} & \colorcelldis{0.35} & \colorcelldis{0.22} & \colorcelldis{0.1} & \colorcelldis{0.1} & \colorcelldis{0.22} & \colorcelldis{0.49} & \colorcelldis{0.35} & \colorcelldis{0.22} & \colorcelldis{0.1} & \colorcelldis{0.1} & \colorcelldis{0.22} \\
\hline
\textbf{Q7} & \colorcelldis{0.7} & \colorcelldis{0.44} & \colorcelldis{0.2} & \colorcelldis{0.83} & \colorcelldis{0.55} & \colorcelldis{0.28} & \colorcelldis{0.0} & \colorcelldis{0.7} & \colorcelldis{0.44} & \colorcelldis{0.2} & \colorcelldis{0.76} & \colorcelldis{0.62} & \colorcelldis{0.49} & \colorcelldis{0.35} & \colorcelldis{0.22} & \colorcelldis{0.1} & \colorcelldis{0.76} & \colorcelldis{0.62} & \colorcelldis{0.49} & \colorcelldis{0.35} & \colorcelldis{0.22} & \colorcelldis{0.1} \\
\hline
\textbf{Q8} & \colorcelldis{0.28} & \colorcelldis{0.39} & \colorcelldis{0.62} & \colorcelldis{0.2} & \colorcelldis{0.2} & \colorcelldis{0.44} & \colorcelldis{0.7} & \colorcelldis{0.0} & \colorcelldis{0.28} & \colorcelldis{0.55} & \colorcelldis{0.22} & \colorcelldis{0.22} & \colorcelldis{0.29} & \colorcelldis{0.4} & \colorcelldis{0.53} & \colorcelldis{0.65} & \colorcelldis{0.1} & \colorcelldis{0.1} & \colorcelldis{0.22} & \colorcelldis{0.35} & \colorcelldis{0.49} & \colorcelldis{0.62} \\
\hline
\textbf{Q9} & \colorcelldis{0.39} & \colorcelldis{0.28} & \colorcelldis{0.39} & \colorcelldis{0.44} & \colorcelldis{0.2} & \colorcelldis{0.2} & \colorcelldis{0.44} & \colorcelldis{0.28} & \colorcelldis{0.0} & \colorcelldis{0.28} & \colorcelldis{0.4} & \colorcelldis{0.29} & \colorcelldis{0.22} & \colorcelldis{0.22} & \colorcelldis{0.29} & \colorcelldis{0.4} & \colorcelldis{0.35} & \colorcelldis{0.22} & \colorcelldis{0.1} & \colorcelldis{0.1} & \colorcelldis{0.22} & \colorcelldis{0.35} \\
\hline
\textbf{Q10} & \colorcelldis{0.62} & \colorcelldis{0.39} & \colorcelldis{0.28} & \colorcelldis{0.7} & \colorcelldis{0.44} & \colorcelldis{0.2} & \colorcelldis{0.2} & \colorcelldis{0.55} & \colorcelldis{0.28} & \colorcelldis{0.0} & \colorcelldis{0.65} & \colorcelldis{0.53} & \colorcelldis{0.4} & \colorcelldis{0.29} & \colorcelldis{0.22} & \colorcelldis{0.22} & \colorcelldis{0.62} & \colorcelldis{0.49} & \colorcelldis{0.35} & \colorcelldis{0.22} & \colorcelldis{0.1} & \colorcelldis{0.1} \\
\hline

\end{tabularx}
\end{table}

\begin{table}[h!]
\caption{Ranked physical distance determining the n-th nearest neighbours }
\label{tab:ranked_distance}
\begin{tabularx}{\textwidth}{|c|X|X|X|X|X|X|X|X|X|X|X|X|X|X|X|X|X|X|X|X|X|X|}
\hline
             & \textbf{P1} & \textbf{P2} & \textbf{P3} & \textbf{P4} & \textbf{P5} & \textbf{P6} & \textbf{P7} & \textbf{P8} & \textbf{P9} & \textbf{P10} & \textbf{B1} & \textbf{B2} & \textbf{B3} & \textbf{B4} & \textbf{B5} & \textbf{B6} & \textbf{B7} & \textbf{B8} & \textbf{B9} & \textbf{B10} & \textbf{B11} & \textbf{B12} \\ \hline
\textbf{Q1}  & \colorcell{1}           & \colorcell{3}           & \colorcell{6}           & \colorcell{2}           & \colorcell{2}           & \colorcell{5}           & \colorcell{8}           & \colorcell{3}           & \colorcell{4}           & \colorcell{7}            & \colorcell{1}           & \colorcell{1}           & \colorcell{2}           & \colorcell{4}           & \colorcell{6}           & \colorcell{8}           & \colorcell{2}           & \colorcell{2}           & \colorcell{3}           & \colorcell{5}            & \colorcell{7}            & \colorcell{9}            \\ \hline
\textbf{Q2}  & \colorcell{3}           & \colorcell{1}           & \colorcell{3}           & \colorcell{5}           & \colorcell{2}           & \colorcell{2}           & \colorcell{5}           & \colorcell{4}           & \colorcell{3}           & \colorcell{4}            & \colorcell{4}           & \colorcell{2}           & \colorcell{1}           & \colorcell{1}           & \colorcell{2}           & \colorcell{4}           & \colorcell{5}           & \colorcell{3}           & \colorcell{2}           & \colorcell{2}            & \colorcell{3}            & \colorcell{5}            \\ \hline
\textbf{Q3}  & \colorcell{6}           & \colorcell{3}           & \colorcell{1}           & \colorcell{8}           & \colorcell{5}           & \colorcell{2}           & \colorcell{2}           & \colorcell{7}           & \colorcell{4}           & \colorcell{3}            & \colorcell{8}           & \colorcell{6}           & \colorcell{4}           & \colorcell{2}           & \colorcell{1}           & \colorcell{1}           & \colorcell{9}           & \colorcell{7}           & \colorcell{5}           & \colorcell{3}            & \colorcell{2}            & \colorcell{2}            \\ \hline
\textbf{Q4}  & \colorcell{2}           & \colorcell{5}           & \colorcell{8}           & \colorcell{1}           & \colorcell{3}           & \colorcell{6}           & \colorcell{9}           & \colorcell{2}           & \colorcell{5}           & \colorcell{8}            & \colorcell{1}           & \colorcell{2}           & \colorcell{4}           & \colorcell{6}           & \colorcell{8}           & \colorcell{10}          & \colorcell{1}           & \colorcell{2}           & \colorcell{4}           & \colorcell{6}            & \colorcell{8}            & \colorcell{10}           \\ \hline
\textbf{Q5}  & \colorcell{2}           & \colorcell{2}           & \colorcell{5}           & \colorcell{3}           & \colorcell{1}           & \colorcell{3}           & \colorcell{6}           & \colorcell{2}           & \colorcell{2}           & \colorcell{5}            & \colorcell{2}           & \colorcell{1}           & \colorcell{1}           & \colorcell{2}           & \colorcell{4}           & \colorcell{6}           & \colorcell{2}           & \colorcell{1}           & \colorcell{1}           & \colorcell{2}            & \colorcell{4}            & \colorcell{6}            \\ \hline
\textbf{Q6}  & \colorcell{5}           & \colorcell{2}           & \colorcell{2}           & \colorcell{6}           & \colorcell{3}           & \colorcell{1}           & \colorcell{3}           & \colorcell{5}           & \colorcell{2}           & \colorcell{2}            & \colorcell{6}           & \colorcell{4}           & \colorcell{2}           & \colorcell{1}           & \colorcell{1}           & \colorcell{2}           & \colorcell{6}           & \colorcell{4}           & \colorcell{2}           & \colorcell{1}            & \colorcell{1}            & \colorcell{2}            \\ \hline
\textbf{Q7}  & \colorcell{8}           & \colorcell{5}           & \colorcell{2}           & \colorcell{9}           & \colorcell{6}           & \colorcell{3}           & \colorcell{1}           & \colorcell{8}           & \colorcell{5}           & \colorcell{2}            & \colorcell{10}          & \colorcell{8}           & \colorcell{6}           & \colorcell{4}           & \colorcell{2}           & \colorcell{1}           & \colorcell{10}          & \colorcell{8}           & \colorcell{6}           & \colorcell{4}            & \colorcell{2}            & \colorcell{1}            \\ \hline
\textbf{Q8}  & \colorcell{3}           & \colorcell{4}           & \colorcell{7}           & \colorcell{2}           & \colorcell{2}           & \colorcell{5}           & \colorcell{8}           & \colorcell{1}           & \colorcell{3}           & \colorcell{6}            & \colorcell{2}           & \colorcell{2}           & \colorcell{3}           & \colorcell{5}           & \colorcell{7}           & \colorcell{9}           & \colorcell{1}           & \colorcell{1}           & \colorcell{2}           & \colorcell{4}            & \colorcell{6}            & \colorcell{8}            \\ \hline
\textbf{Q9}  & \colorcell{4}           & \colorcell{3}           & \colorcell{4}           & \colorcell{5}           & \colorcell{2}           & \colorcell{2}           & \colorcell{5}           & \colorcell{3}           & \colorcell{1}           & \colorcell{3}            & \colorcell{5}           & \colorcell{3}           & \colorcell{2}           & \colorcell{2}           & \colorcell{3}           & \colorcell{5}           & \colorcell{4}           & \colorcell{2}           & \colorcell{1}           & \colorcell{1}            & \colorcell{2}            & \colorcell{4}            \\ \hline
\textbf{Q10} & \colorcell{7}           & \colorcell{4}           & \colorcell{3}           & \colorcell{8}           & \colorcell{5}           & \colorcell{2}           & \colorcell{2}           & \colorcell{6}           & \colorcell{3}           & \colorcell{1}            & \colorcell{9}           & \colorcell{7}           & \colorcell{5}           & \colorcell{3}           & \colorcell{2}           & \colorcell{2}           & \colorcell{8}           & \colorcell{6}           & \colorcell{4}           & \colorcell{2}            & \colorcell{1}            & \colorcell{1}            \\ \hline
\end{tabularx}
\end{table}

\newpage

\section{Dependence of g-factor on the hole occupancy}

\begin{figure*}[htb!]
	\centering
	\includegraphics{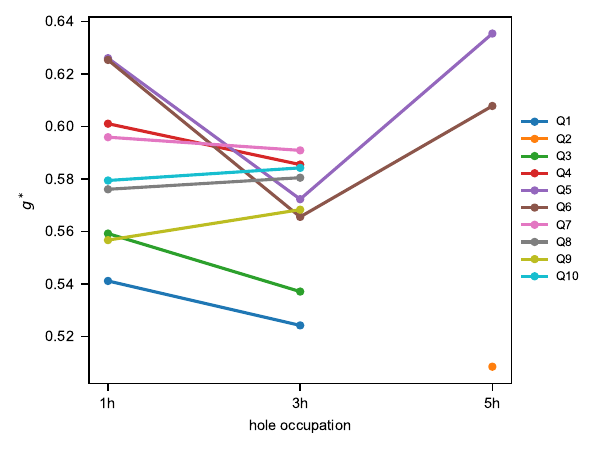}
	\caption{Effective g-factor values of the 10 qubits in the single, three and five-hole occupation.}
	\label{fig:gfactor_vs_occupation_line_plot}
\end{figure*}

\clearpage
\section{EDSR driving efficiency}

\begin{figure*}[htb!]
	\centering
	\includegraphics{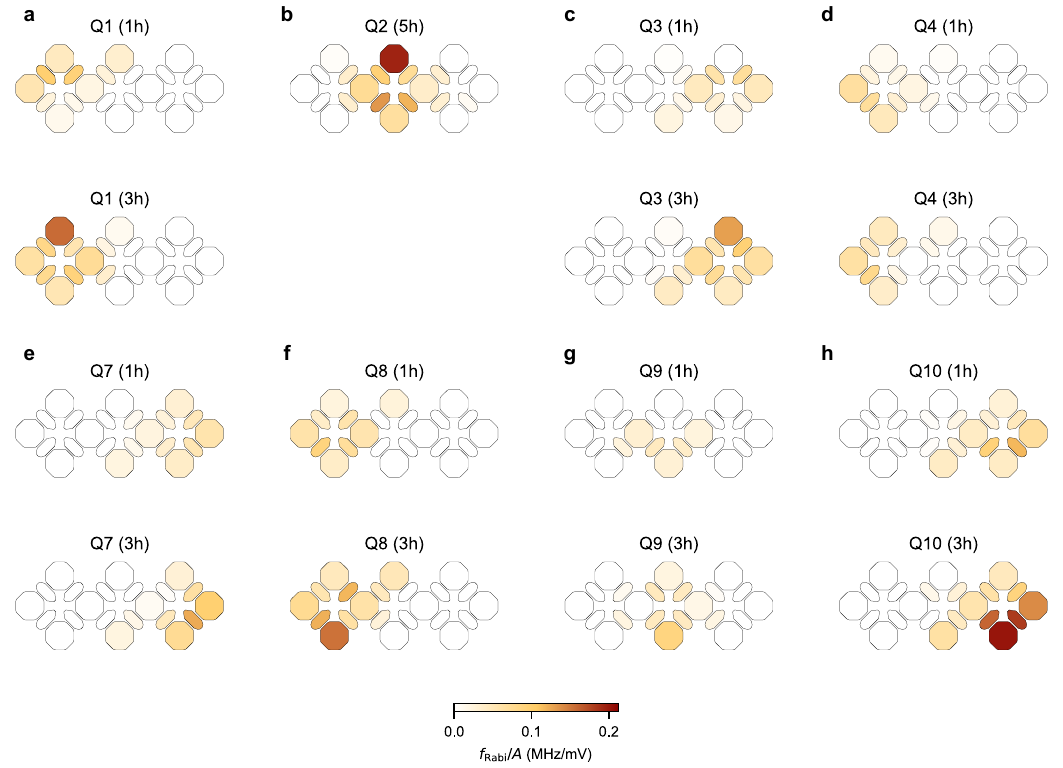}
	\caption{Extended data of the EDSR driving efficiency for Q1 (a), Q2 (b), Q3 (c), Q4 (d), Q7 (e), Q8 (f), Q9 (g), Q10 (h) in the single- and three- hole configuration for all qubits except Q2 which is only probed in the five-hole occupancy. Note that Q4 and Q7 show less prominent increase of the top plunger drive efficiency, which could be attributed to their large charging voltage as discussed in the main text.}
	\label{fig:driving_eff}
\end{figure*}

Additional data of EDSR driving efficiency as a function of driving gate and charge occupation are shown in \autoref{fig:driving_eff}.

\clearpage
\section{g-factor tunability}

\begin{figure*}[htb!]
	\centering
	\includegraphics{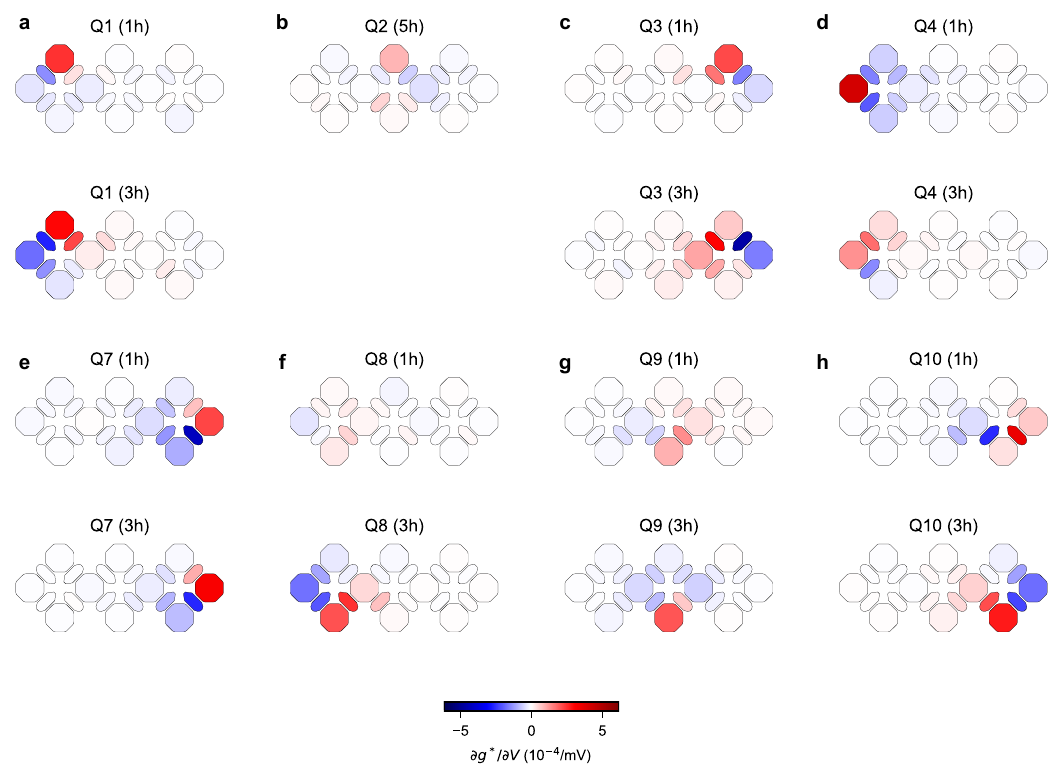}
	\caption{Extended data of the g-factor tunability for Q1 (a), Q2 (b), Q3 (c), Q4 (d), Q7 (e), Q8 (f), Q9 (g), Q10 (h) in the single- and three- hole configuration for all qubits except Q2 which is only probed in the five-hole occupancy. More details about data analysis can be found in Suppl. Note S10.}
	\label{fig:gfactor_tunability}
\end{figure*}

\clearpage

\section{LSES extraction}

\begin{figure*}[h!]
	\centering
	\includegraphics[width=14cm]{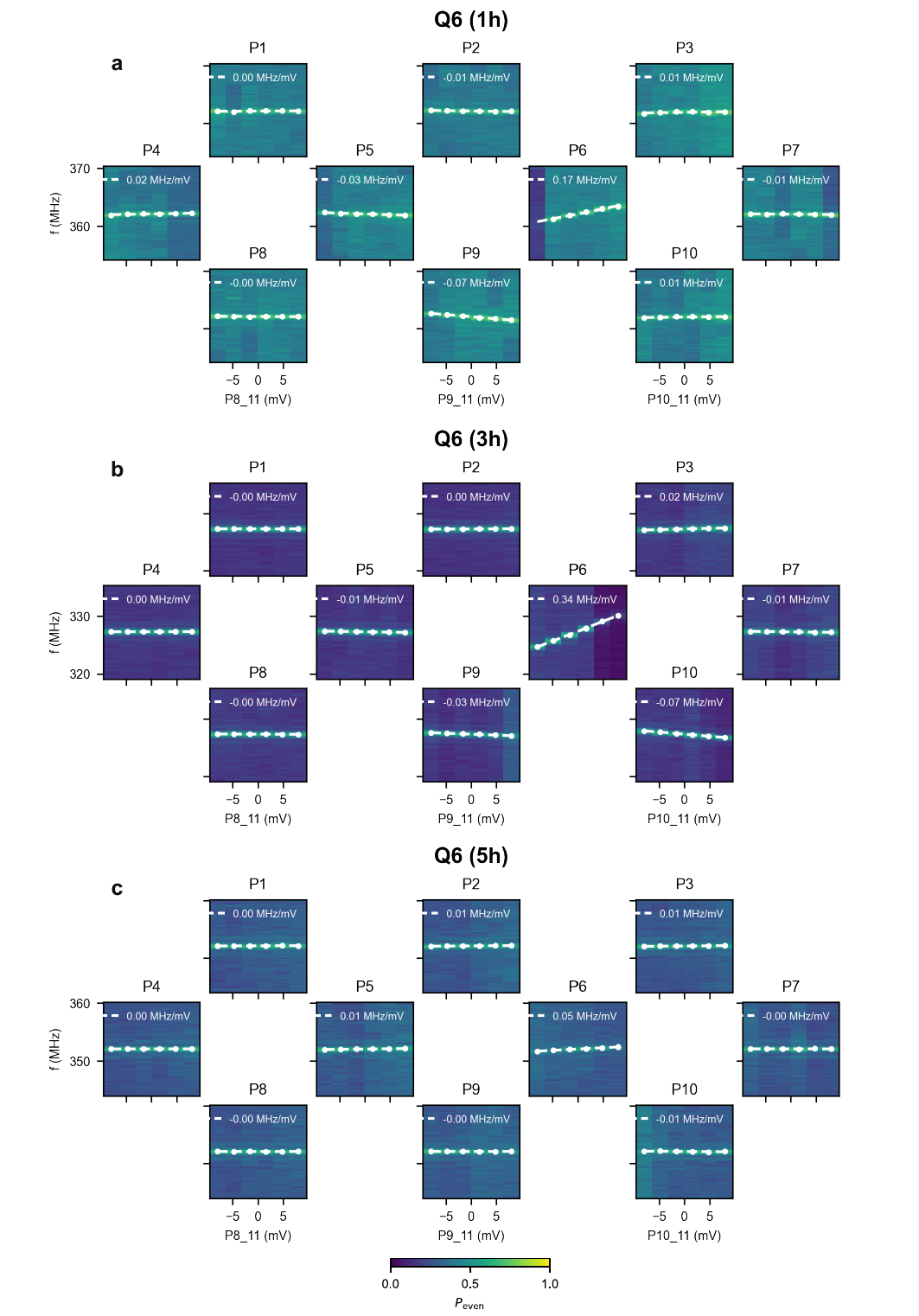}
	\caption{\textbf{LSES contribution of plungers for Q6 with 1, 3, and 5 hole occupation.} 
	}
	\label{fig:LSES_Q6_plunger}
\end{figure*}

\begin{figure*}[h!]
	\centering
	\includegraphics[width=14cm]{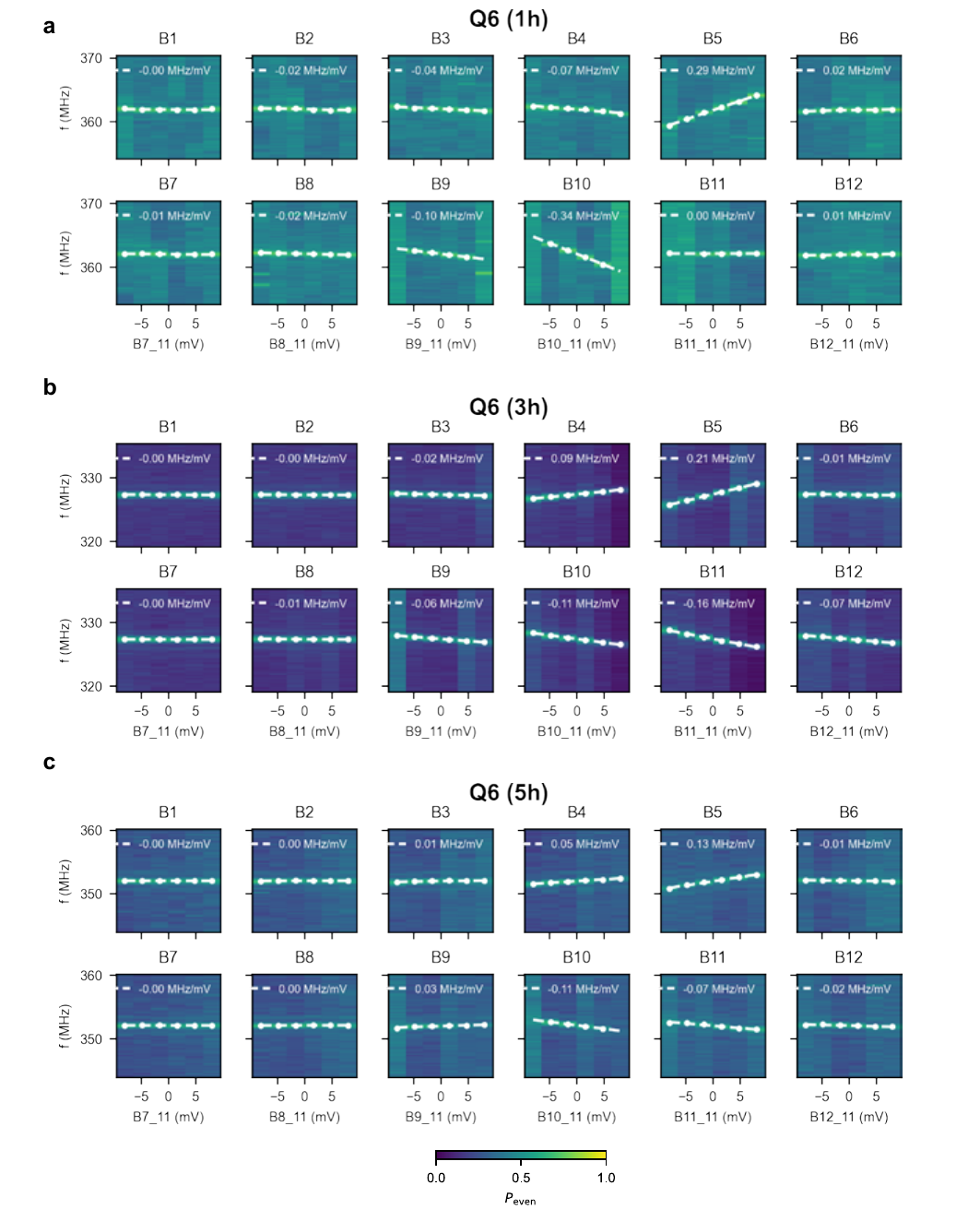}
	\caption{\textbf{LSES contribution of barriers for Q6 with 1, 3, and 5 hole occupation.} 
	}
	\label{fig:LSES_Q6_barriers}
\end{figure*}

To extract the g-factor susceptibility of a qubit to each gate, microwave frequency sweeps around the larmor frequency have been performed, while changing the applied voltage on each gate one after the other. In this experiment, we used a chirp signal to probe the resonance frequency of each qubit. After, the resonance frequency for each configuration is determined and fitted with a linear fit. The corresponding slope indicates the gate susceptibility $\partial f_{\mathrm{Rabi}} / \partial V_{\mathrm{gate}}$. In \autoref{fig:LSES_Q6_plunger} and \ref{fig:LSES_Q6_barriers} examples of the data and the corresponding linear fits are plotted.

\newpage

\clearpage

\section{Driving efficiency extraction}

\begin{figure*}[h!]
	\centering
	\includegraphics[width=14cm]{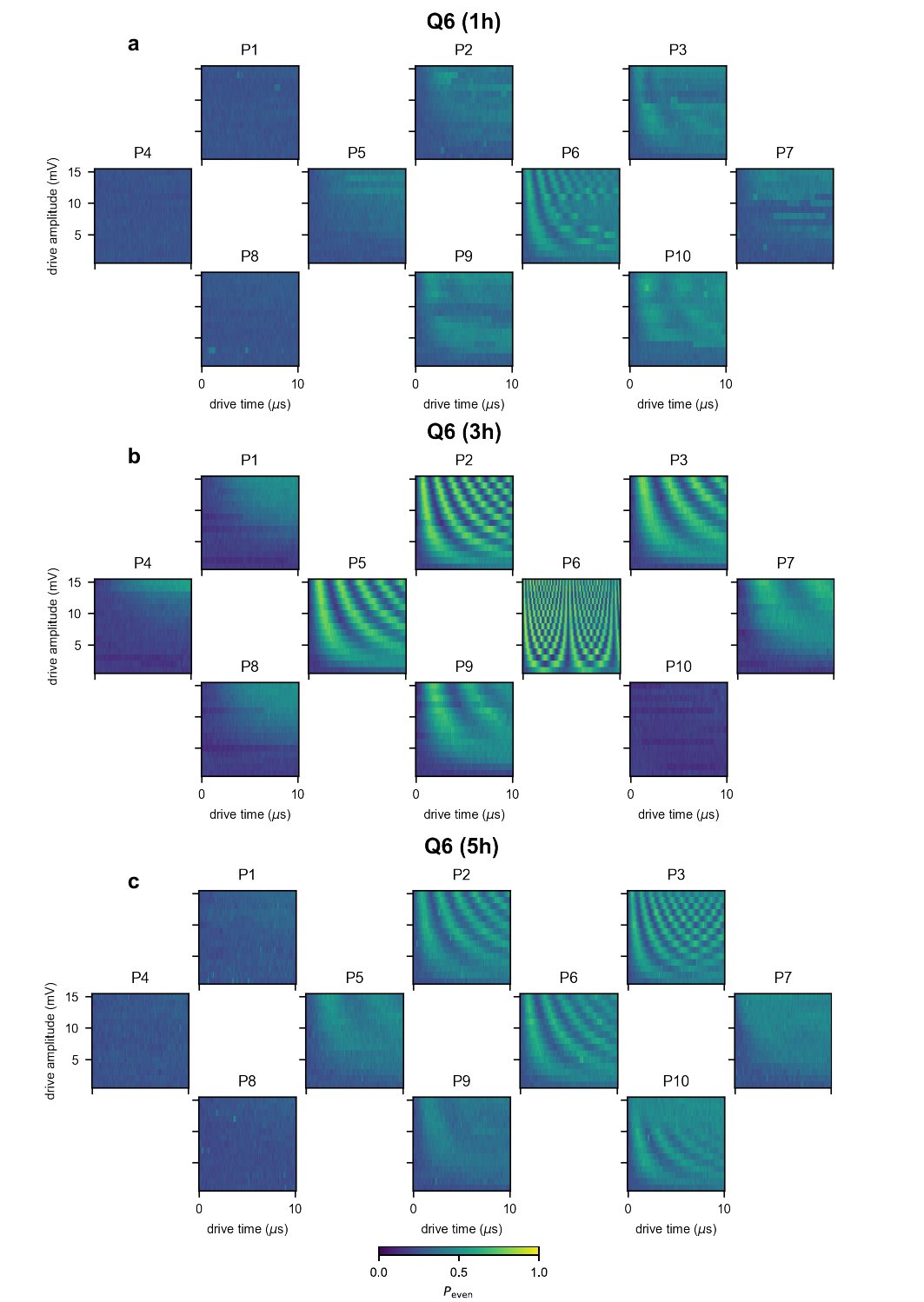}
	\caption{\textbf{EDSR driving of plungers for Q6 with 1, 3, and 5 hole occupation.} 
	}
	\label{fig:Eff_raw_Q6_plunger}
\end{figure*}

\begin{figure*}[h!]
	\centering
	\includegraphics[width=14cm]{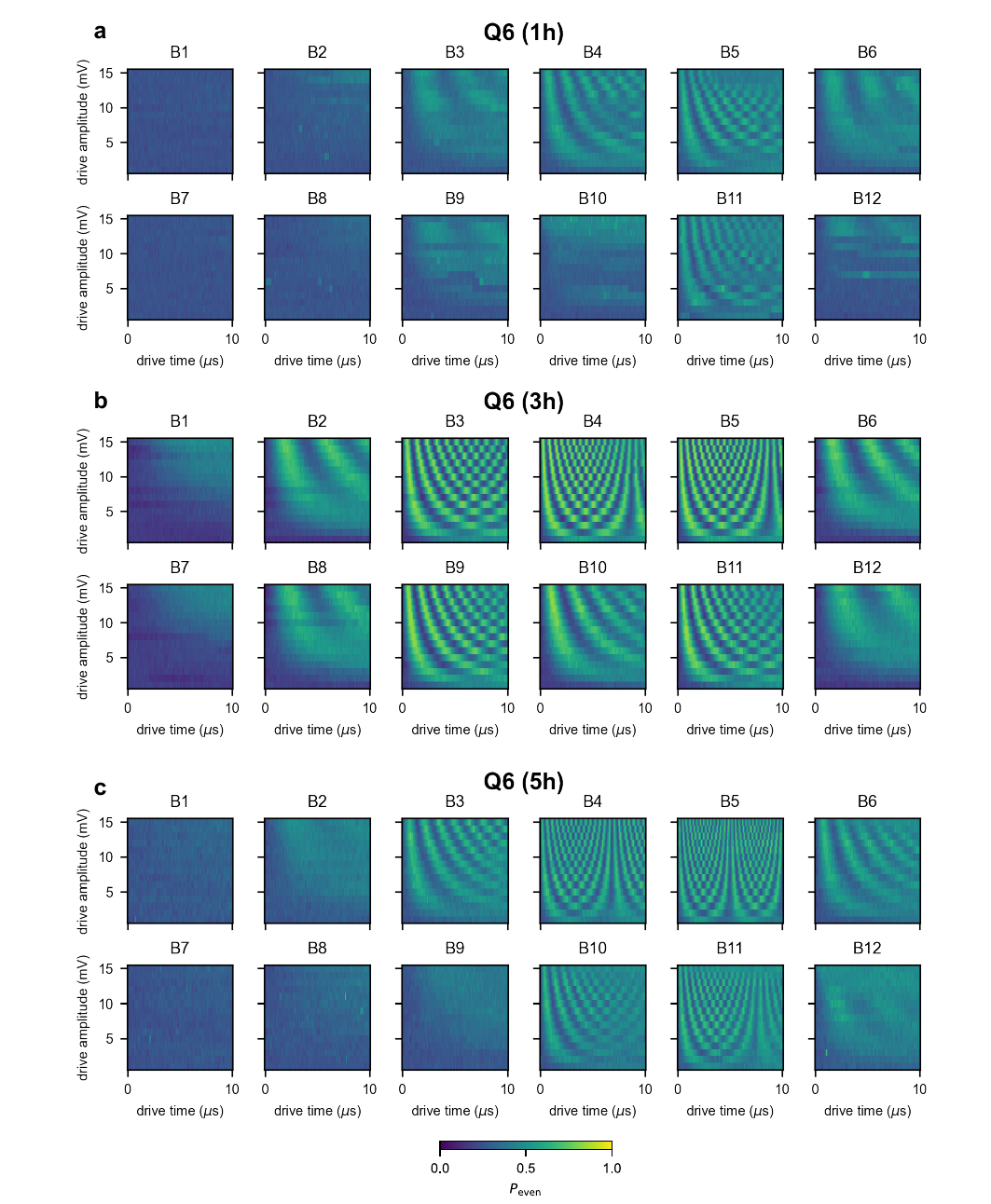}
	\caption{\textbf{EDSR driving of barriers for Q6 with 1, 3, and 5 hole occupation.} 
	}
	\label{fig:Eff_raw_Q6_barriers}
\end{figure*}

\begin{figure*}[h!]
	\centering
	\includegraphics[width=14cm]{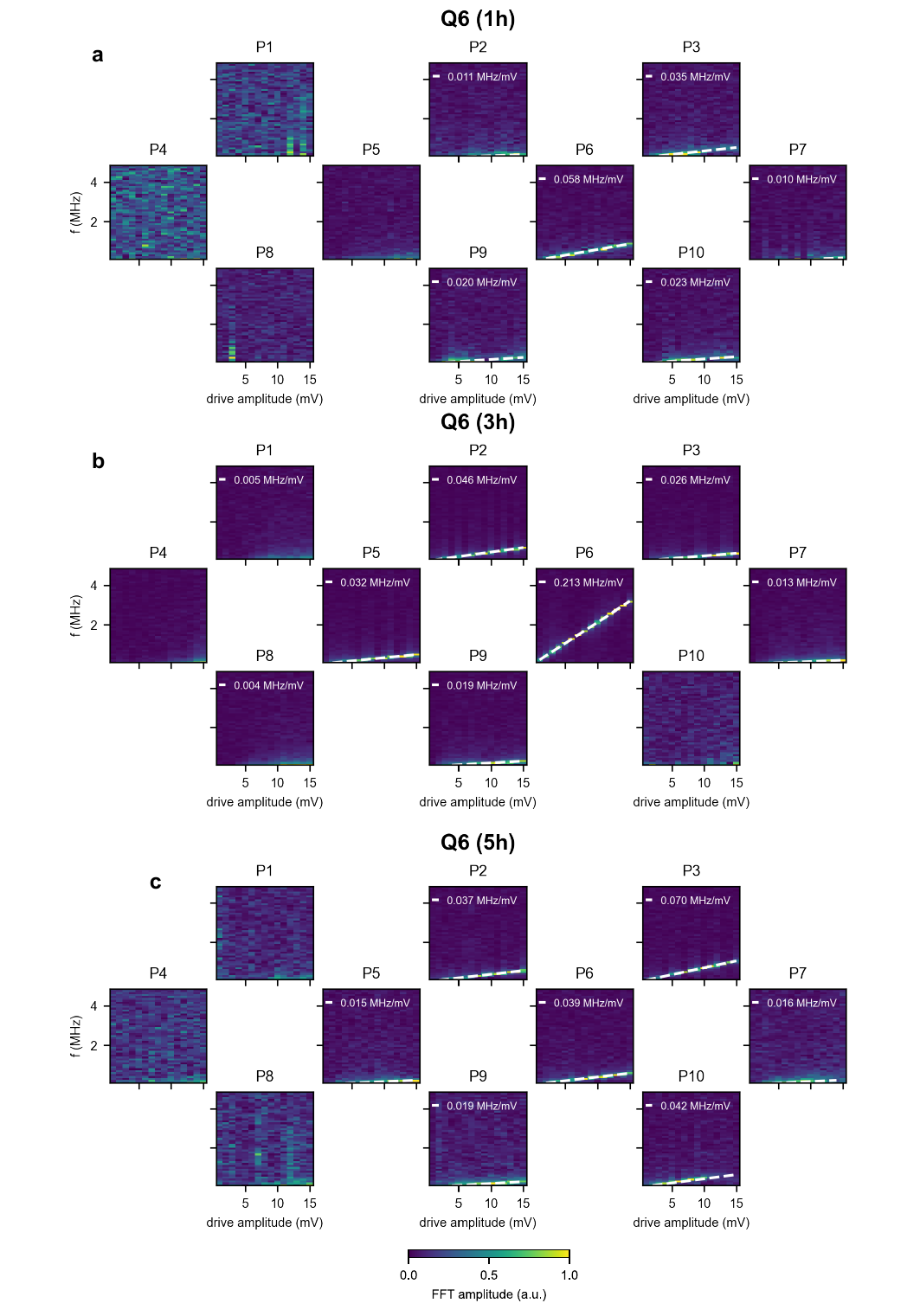}
	\caption{\textbf{FFT and linear fit of EDSR driving of plungers for Q6 with 1, 3, and 5 hole occupation.} 
	}
	\label{fig:Eff_raw_fft_Q6_plunger}
\end{figure*}

\begin{figure*}[h!]
	\centering
	\includegraphics[width=14cm]{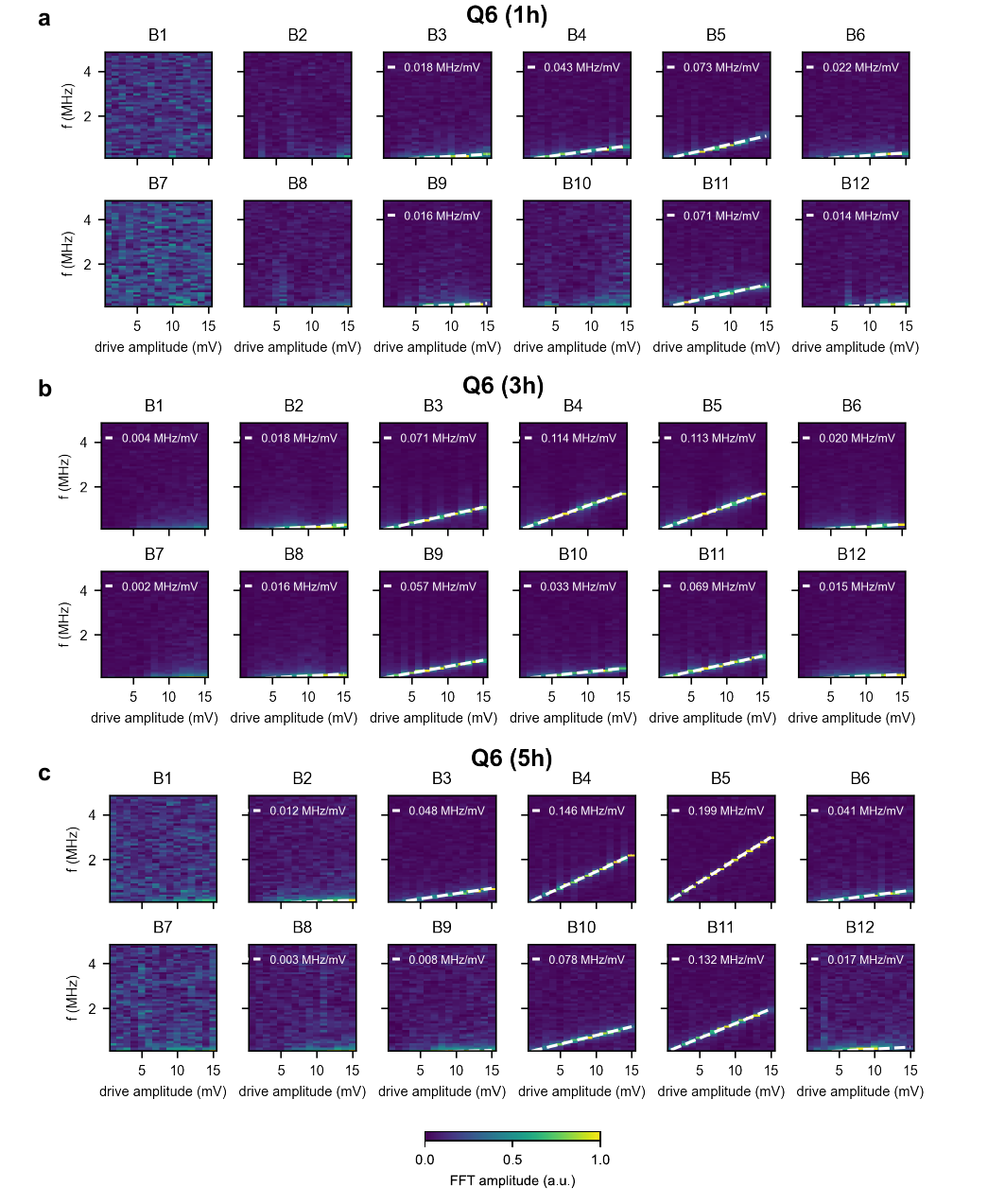}
	\caption{\textbf{FFT and linear fit of EDSR driving of barriers for Q6 with 1, 3, and 5 hole occupation.} 
	}
	\label{fig:Eff_raw_fft_Q6_barriers}
\end{figure*}

To extract the EDSR driving efficiency, Rabi measurements are performed as a function of drive amplitude. Next a fast-Fourier transform is applied to the raw data, and then fitted. In Suppl. Figs. \ref{fig:Eff_raw_Q6_plunger}-\ref{fig:Eff_raw_fft_Q6_barriers} examples of the data and the corresponding linear fits are plotted. The slope determines the driving efficiency.

\newpage

\newpage

\newpage

\clearpage

\section{Modelling of single- and multi-hole quantum dots}

We have modelled the devices analytically and numerically to understand the trends highlighted by the experiments. We first introduce the structural models used in the simulations, then discuss the results for single-hole quantum dots. We finally derive an analytical model for the Rabi frequencies of multiply charged dots. This model, backed by numerical simulations, supports the enhancement of the Rabi frequencies in three-hole quantum dots.

\subsection{Models and device}

\begin{figure}[t]
\centering
\includegraphics[width=8 cm]{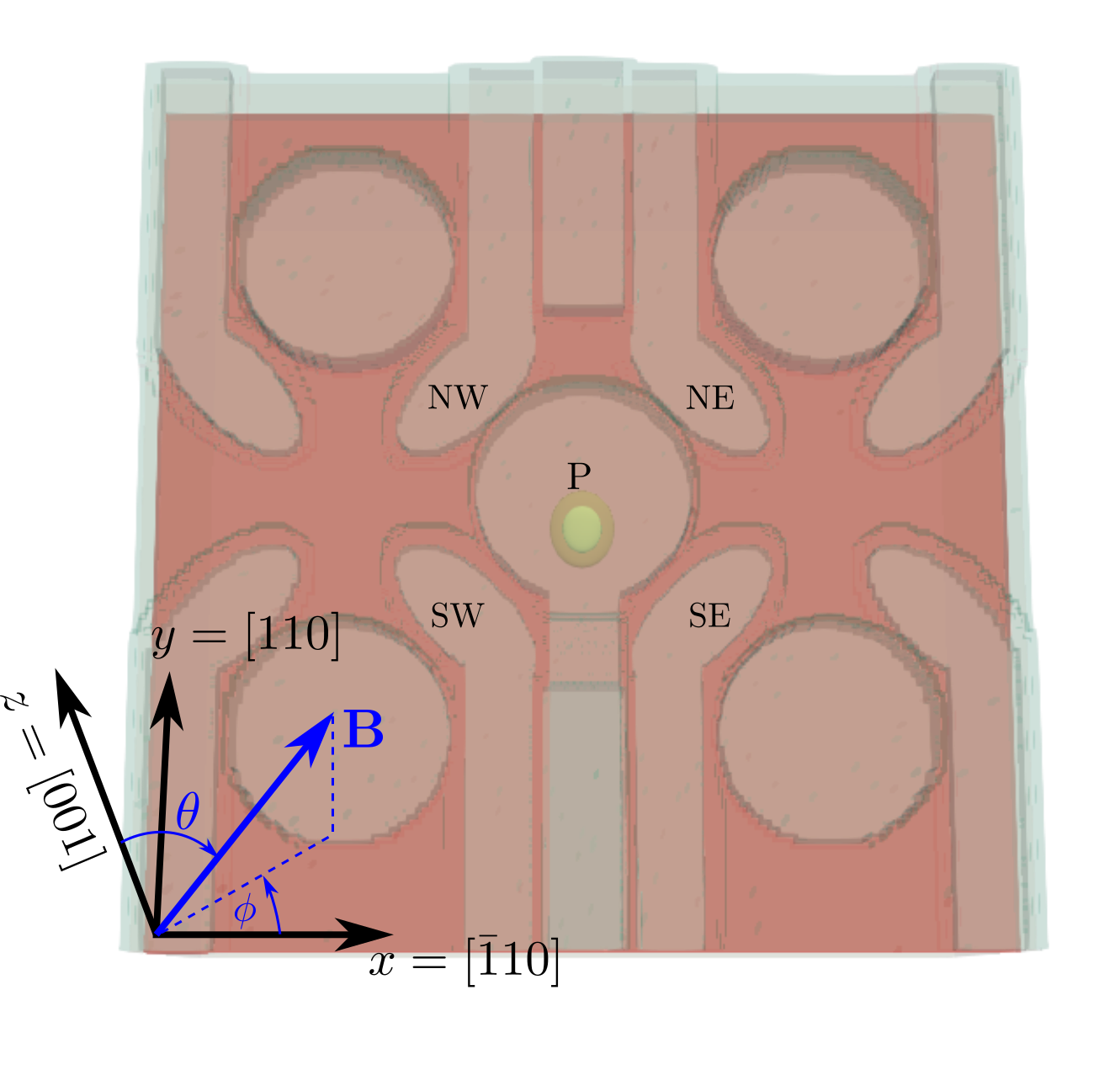}
\caption{The simulated device geometry, which includes the three layers of gates (barrier gates NW/NE/SW/SE, screening gates, and plunger gate P). The yellow shape below the plunger gate is the iso-density surface that encloses 80\% of the charge of the ``centered'' dot of Fig.~\ref{fig:single}a-c. The orientation of the magnetic field $\vec{B}$ is characterized by the angles $\theta$ and $\phi$ in the crystallographic axes set $x=[\bar{1}10]$, $y=[110]$ and $z=[001]$.}
\label{fig:device}
\end{figure}

We consider the simplified geometry of Fig.~\ref{fig:device}, comprising a central plunger gate separated from its four nearest neighbors by the north-west (NW), north-east (NE), south-east (SE), and south-west (SW) barrier gates. Screening gates are also included between the NW/NE and SW/SE gate lines, to prevent, in particular, accumulation below the plunger gate line. The heterostructure and gate stack as well as the dimensions of the gates are borrowed from the experimental layout. This model geometry is close to the environment of dots Q5 and Q6.

We compute the potential in the heterostructure with a finite-volume Poisson solver, then the single-hole wave functions with finite-differences implementation of the four-band Luttinger-Kohn model~\cite{Luttinger1955MotionFields,Willatzen2009TheSemiconductors}, and finally the LSES and Rabi frequencies with the $g$-matrix formalism \cite{Venitucci2018ElectricalFormalism,Martinez2022HoleFields,Mauro2024DephasingSweetSpot}. We also compute the three-hole ground state and g-matrix with a full configuration interaction (FCI) method \cite{Abadillo-Uriel2021Two-bodyQubits}. Moreover, we construct a phenomenological theory that qualitatively captures the main features observed in the experiment by including only the few most relevant configurations of this FCI model.

We add fixed trapped charges with density $\sigma=5\times 10^{11}$\,$e$/cm$^2$ at the semiconductor/Al$_2$O$_3$ interface, either as a homogeneous sheet (which does not introduce disorder), or as a random distribution of point charges (see discussion below). We do not account here for the inhomogeneous strains imposed by the contraction of the metal gates upon cool-down \cite{Abadillo-Uriel2023Hole-SpinInteractions}. The results with such inhomogeneous strains are qualitatively similar; the Rabi frequencies with and without cool-down strains are broadly comparable at the experimental magnetic field orientation, while the average LSES of the barrier gates is typically smaller (resp. larger) than the experiment without (resp. with) these strains. This suggests that only part of the strains have been transferred to the heterostructure (due to, e.g., plasticity at the metal/oxide or oxide/semiconductor interfaces).

We first discuss the conclusions drawn from the modeling of single-hole dots, then of three-hole dots.

\subsection{Single-hole dots}
\label{sec:SP}

\begin{figure}[t]
\centering
\includegraphics[width=0.8\columnwidth]{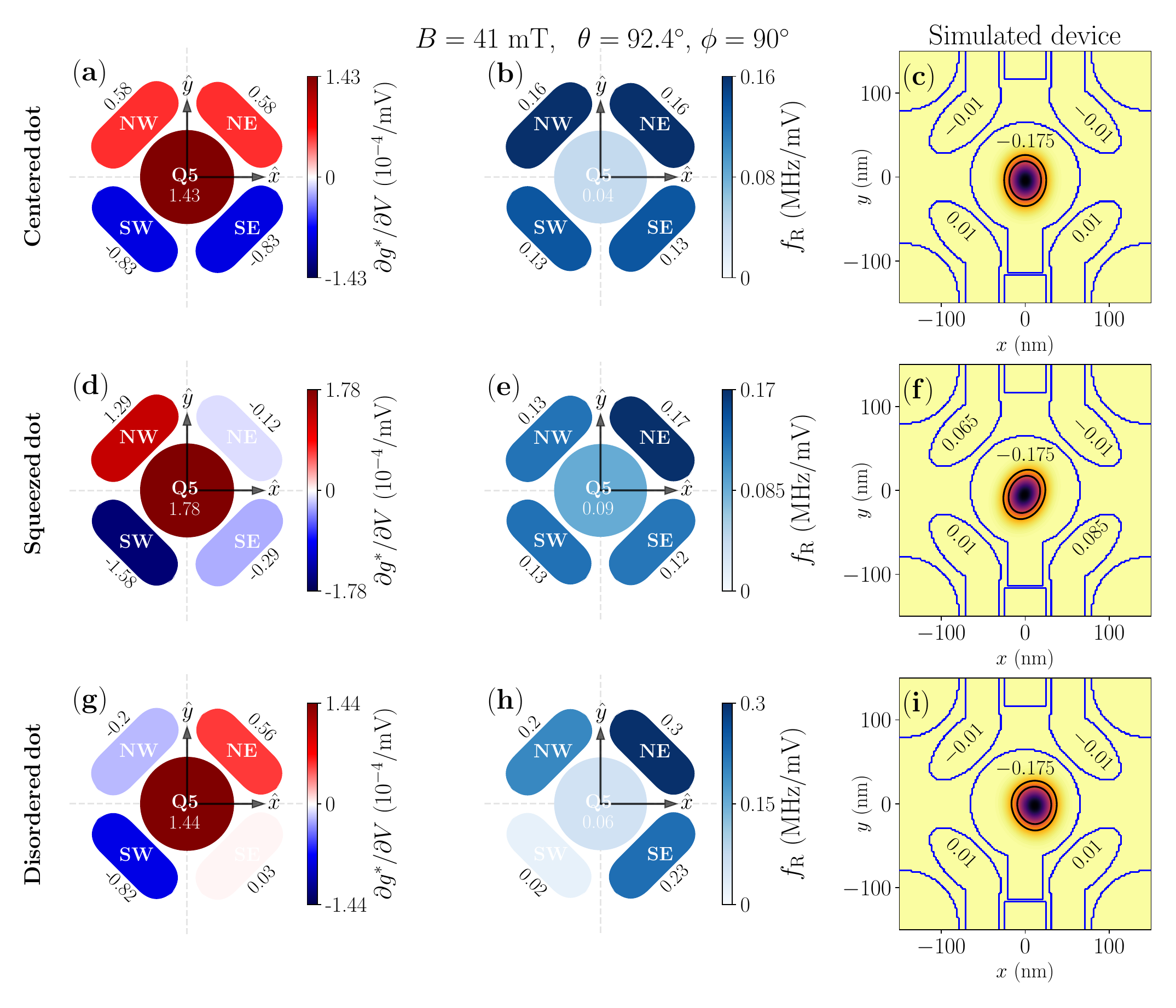}
\caption{(a-c) LSES, Rabi frequencies $f_\mathrm{R}$ and map of the squared wave function computed for a single hole in a ``centered'' dot. The LSES $\partial g/\partial V$ of the plunger and barrier gates (per mV) are reported in panel a), while the Rabi frequencies (in MHz/mV) are reported in panel b), and the bias voltages (in V) are reported in panel c). (d-f) Same for a ``squeezed'' dot at a different bias point. (g-i) Same for the displaced dot with charge disorder at the semiconductor/Al$_2$O$_3$ interface.}
\label{fig:single}
\end{figure}

We start from a bias point (Fig.~\ref{fig:single}a-c) where the ground-state hole wave function is well centered within the dot. The difference of potentials between the plunger and barrier gates ($\simeq 175$\,mV) is similar to the experiment. We do not account for disorder at this stage (the distribution of charges at the semiconductor/Al$_2$O$_3$ interface is homogeneous). The dot is nonetheless slightly squeezed due to the asymmetry of the structure.

The g-tensor of the hole can be diagonalized in order to identify the principal g-factors $g_X$, $g_Y$ and $g_Z$ and the gyro-magnetic axes $X$, $Y$, $Z$; in this axis set, the effective g-factor $g^*$ simply reads
\begin{equation}
g^*=\sqrt{g_X^2 b_X^2+g_Y^2 b_Y^2+g_Z^2 b_Z^2}\,,   
\label{eq:gstar}
\end{equation}
where $(b_X,\,b_Y,\,b_Z)$ are the coordinates of the unit vector oriented along the magnetic field $\vec{B}$ \cite{Abragam1971ElectronIons,Venitucci2018ElectricalFormalism}. For a perfectly centered and symmetric dot, the gyro-magnetic axes coincide with the device $x$, $y$, $z$ axes; in the present case $g_x^*\equiv|g_X|=0.16$, $g_y^*\equiv|g_Y|=0.043$ and $ g_z^*\equiv|g_Z|=13.46$. This strong anisotropy between in-plane and out-of-plane g-factors is characteristic of heavy-holes in Germanium. The imbalance between $g_X$ and $g_Y$ results from the slight squeezing of the dot \cite{Bosco2021SqueezedPower,Mauro2024DephasingSweetSpot}. For a magnetic field $\vec{B}=B(0,\,\sin\theta,\,\cos\theta)$ oriented $2.4^\circ$ out of plane ($\theta=92.4^\circ$), the calculated effective g-factor $g^*=0.56$ is dominated by the out-of-plane component $g_Z$.

The magnitudes of the LSES and Rabi frequencies computed at $B=41$\,mT are reasonably comparable to the experiment given the uncertainties of the model (strains, exact nature and distribution of traps, Coulomb interactions between neighboring dots, ....). The LSES show generic features weakly dependent on the bias point and disorder:
\begin{itemize}
    \item The LSES of the plunger gate is always positive and usually comparable or larger than the magnitude of the LSES of the barrier gates.
    \item The LSES of the barrier gates alternate positive and negative signs.
\end{itemize}
Indeed, raising the plunger gate voltage primarily deconfines the dot, which results in an increase of both in-plane and out-of-plane g-factors, thus in a strongly positive LSES \cite{Mauro2024DephasingSweetSpot}. On the opposite, raising a barrier gate voltage further confines but deforms the dot, which results, in particular, in a decrease of $g_Z$. For such a symmetric dot, the LSES of all four barrier gates would actually be negative (blue) if these gates were only acting on the principal g-factors $g_X$, $g_Y$ and $g_Z$. However, the barrier gates also tilt the gyro-magnetic $X$, $Y$, $Z$ axes owing to the coupling between the in-plane and out-of-plane motions of the hole in the non-separable confinement potential of the dot (and owing to the inhomogeneous cool-down strains, when present) \cite{Martinez2022HoleFields, Abadillo-Uriel2023Hole-SpinInteractions}. In particular, the gyro-magnetic $Z$ axis rotates by a small angle $\delta\theta<0$ when raising the SW and SE gate voltages. This brings the magnetic field closer to the effective equatorial $(XY)$ plane, which decreases the contribution from the out-of-plane g-factor $g_Z$ in Eq.~\eqref{eq:gstar} and further reduces the net $g^*$. On the opposite, raising the NW and NE gate voltages brings the magnetic field farther from the effective equatorial plane, which overcomes the decrease of $g_Z$ and gives rise to a positive LSES.

The Rabi oscillations essentially result from the modulations of the principal g-factors $g_X$, $g_Y$ and $g_Z$ by the driving gate (g-tensor modulation resonance or g-TMR) \footnote{The tilt of the gyromagnetic axes induced by the drive (owing to the coupling between the in-plane and vertical motions or to inhomogeneous strains) makes little contribution to the Rabi frequencies here; Indeed, it gives rise to a $\propto\sigma_z$ drive term in the Hamiltonian \cite{Martinez2022HoleFields, Abadillo-Uriel2023Hole-SpinInteractions}, which is ``transverse'' (inducing Rabi oscillations) when the magnetic field is in-plane, but becomes longitudinal (enhancing LSES) once $\vec{B}$ goes out-of-plane and the heavy-hole spin gets locked onto the $z$ axis.}. The plunger gate is actually expected to be inefficient when the magnetic field is strictly in-plane \cite{Martinez2022HoleFields, Abadillo-Uriel2023Hole-SpinInteractions}. Indeed, a disk-shaped quantum dot breathes homogeneously in the radio-frequency electric field of the plunger gate, which identically modulates $|g_X|$ and $|g_Y|$, and therefore does not act on the spin precession axis. This gives rise to a large LSES (as highlighted above), but to no transverse coupling (Rabi oscillations). The efficiency of the plunger gate however increases when the dot is significantly squeezed (because breathing is not isotropic in the $XY$ plane any more) and/or when the magnetic field goes out of plane. 

The strength of these spin-orbit coupling mechanisms depends on the symmetry of the dot. The position and shape of the hole wave function can, in particular, be controlled by the barrier gates voltages. This is illustrated in Fig.~\ref{fig:single}d-f, where the bias has been tuned to squeeze the dot along the SW-NE axis. The LSES of three out of the four barrier gates are now negative as a result of the new imbalance between the variations of the principal g-factors and the rotations of the gyro-magnetic axes (around $x$, $y$ and $z$). Yet the LSES of the plunger gate remains positive (whatever the position of the dot). Indeed, the dot is still essentially breathing when raising the plunger gate voltage, which increases all principal g-factors but hardly rotates the gyro-magnetic axes. The balance between the Rabi frequencies of the plunger and barrier gates is also impacted by the stronger asymmetry; in particular, the efficiency of the plunger gate is now comparable to the efficiency of the barrier gates (see above discussion).

Disorder can also change the symmetry of the hole wave function \cite{Martinez2024MitigatingLayout}. It turns out that disorder, even weak, has a strong impact on the sign of the LSES of the barrier gates, owing to the presence of ``sweet lines'' (zero LSES) of these gates near the equatorial plane of the unit sphere describing the magnetic field orientation \cite{Mauro2024DephasingSweetSpot}. The disorder may shuffle these sweet lines so that the magnetic field can practically end up on either side (positive or negative LSES). The effects of disorder are illustrated in Fig.~\ref{fig:single}g-i, for a particular distribution of positive point charges $(\sigma=5\times 10^{11}$\,$e$/cm$^{-2}$) at the semiconductor/Al$_2$O$_3$ interface. We emphasize that this disorder is actually weak and has little incidence on the position and shape of the hole wave function (same bias point as in Fig.~\ref{fig:single}a-c). It has, nonetheless, significant impact on their derivatives (thus on the sign of the LSES and magnitude of the Rabi frequencies). Indeed, disorder does not only control the symmetry of the dot (together with the bias voltages); it also changes the response of the hole to electrical perturbations (as it ``pins'' the motion of the dot to some extent). This particular realization of disorder is in qualitative agreement with the experimental data for Q6 (color pattern of the LSES and Rabi frequencies); we can not claim however that Fig.~\ref{fig:single}i is a fair representation of the wave function of the singly-occupied Q6 as the matching bias/disorder is not unique for a single magnetic field orientation.

\subsection{Three-hole dots}

We now address the three-hole case, starting with a discussion of the effects of Coulomb interactions on  Rabi oscillations, next illustrated with FCI calculations. 

The multi-spin system of many particles is exactly described by the Hamiltonian~\cite{Burkard2023SemiconductorQubits}
\begin{align}
    H=&\sum_{\alpha} \left( \epsilon_\alpha \sigma_0 + \mu_B \boldsymbol{\sigma}\cdot g_{\alpha}\mathbf{B}\right) \nonumber\\
      &+\sum_{\alpha\neq\beta} \tau^{s_\alpha,s_\beta}_{\alpha,\beta} c^\dagger_{\alpha,s_\alpha}c_{\beta,s_\beta} \nonumber\\
      &+\sum_{\alpha,\beta,\gamma,\delta} \Gamma^{s_\alpha,s_\beta,s_\gamma,s_\delta}_{\alpha,\beta,\gamma,\delta} c^\dagger_{\alpha,s_\alpha}c_{\beta,s_\beta}c^\dagger_{\gamma,s_\gamma}c_{\delta,s_\delta}\,,
\end{align}
where $\sigma_0$ is the $2\times2$ identity matrix, and $\boldsymbol{\sigma}=(\sigma_x,\sigma_y,\sigma_z)^T$ is the Pauli-vector consisting of the three Pauli matrices. $\tau_{\alpha,\beta}$ are the standard inter-orbital tunneling elements between orbital $\alpha$ and $\beta$ with spin $s_\alpha,s_\beta=\,\uparrow,\downarrow$ and $\Gamma_{\alpha,\beta,\gamma,\delta}$ are the Coulomb matrix elements connecting $\alpha,\beta,\gamma,\delta$ orbitals with spin $s_\alpha,s_\beta,s_\gamma,s_\delta=\,\uparrow,\downarrow$. The associated many-body wavefunctions can for example be constructed from single-particle eigenstates as described in Ref.~\cite{Burkard2023SemiconductorQubits}. In that case, $\tau_{\alpha,\beta} = 0$.

Consequently, our system of interest, a single dot filled by 3 holes and orbitals $\alpha,\beta,\gamma,\delta=\lbrace \ket{0},\ket{1},\cdots, \ket{k} \rbrace$ is described by
\begin{align}
    H=\begin{pmatrix}
    \boldsymbol{\sigma}\cdot g_{E_0}\mathbf{B} + \sigma_0 \mathcal{E}_{\text{orb},E_0} && \sigma_0 t_{E_0,E_1} && \cdots && \sigma_0 t_{E_0,E_k} \\
    \sigma_0 t_{E_1,E_0} && \boldsymbol{\sigma}\cdot g_{E_1}\mathbf{B} + \sigma_0 \mathcal{E}_{\text{orb},E_1} && \cdots  && \vdots \\
    \vdots && \vdots && \ddots && \vdots\\
    \sigma_0 t_{E_k,E_0} && \cdots && \cdots && \boldsymbol{\sigma}\cdot g_{E_k}\mathbf{B} + \sigma_0 \mathcal{E}_{\text{orb},E_k}
    \end{pmatrix}\,.
    \label{eq:FCImatrix}
\end{align}
Here, $\ket{E_0}$ describes a configuration where the ground state orbital $\ket{0}$ is doubly occupied and the remaining hole occupies the lowest excited orbital $\ket{1}$. The states $\ket{E_k}$ with $k>0$ are excited configurations. Considering only single excitations, the tunnel matrix elements are given by $t_{\alpha,\beta} = \tau_{\alpha,\beta} + \sum_{\gamma}( \Gamma_{\alpha,\beta,\gamma,\gamma} + \Gamma_{\gamma,\gamma,\alpha,\beta} + \Gamma_{\alpha,\gamma,\gamma,\beta} + \Gamma_{\gamma,\beta,\alpha,\gamma})$, where the sum $\gamma$ goes over all occupied orbitals. Similarly, the orbital energies are given by $\mathcal{E}_{\text{orb},\alpha}=\epsilon_\alpha + \sum_\gamma (\epsilon_\gamma + \Gamma_{\alpha,\alpha,\gamma,\gamma} + \Gamma_{\gamma,\gamma,\alpha,\alpha} + \Gamma_{\alpha,\gamma,\gamma,\alpha} + \Gamma_{\gamma,\alpha,\alpha,\gamma})$.
Additionally, we neglect the spin-orbit interactions, which are usually small in Germanium, especially at the intra-dot scale. However, in general, one can add the spin-orbit contributions to all inter-orbital transition matrix elements $\sigma_0 t_{\alpha,\beta}\rightarrow \cos(\zeta_{\alpha,\beta})\sigma_0 t_{\alpha,\beta} + \sin(\zeta_{\alpha,\beta})\boldsymbol{\sigma}\cdot \mathbf{t}_{\text{soi},\alpha,\beta}$ and orbital energies $\sigma_0 \mathcal{E}_{\text{orb},\alpha}\rightarrow \cos(\zeta_{\alpha})\sigma_0 \mathcal{E}_{\text{orb},\alpha} + \sin(\zeta_{\alpha})\boldsymbol{\sigma}\cdot \mathbf{\mathcal{E}}_{\text{orb,soi},\alpha}$. 
In the main text, we focused on the weak coupling strength case $|t_{E_0,E_k}|/(\mathcal{E}_{\text{orb},E_k}-\mathcal{E}_{\text{orb},E_0})|\ll 1$ using perturbation theory. While this captures the main experimental features, such as the strongly altered Rabi frequencies with respect to the single hole case, we show below that our model holds more generally.

\subsubsection{Analytical expressions}
We now consider the strong coupling case. For simplicity, we assume $|t_{E_0,E_1}|/(\mathcal{E}_{\text{orb},E_1}-\mathcal{E}_{\text{orb},E_0})\gg|t_{E_0,E_k}|/(\mathcal{E}_{\text{orb},E_k}-\mathcal{E}_{\text{orb},E_0})|$, which is expected for two-fold quasi-degeneracy, e.g. $p$-orbitals of the 2D harmonic oscillator. Consequently, we focus solely on the space spanned by $\lbrace\ket{E_0},\ket{E_1}\rbrace$. This simplification allows us to construct a simple yet meaningful analytical theory, which is analogous to that of a flopping mode spin qubit. With this theory, we can qualitatively interpret many experimental features. We note, that the case of multiple strongly coupled orbitals can be treated similarly. To be more quantitative, we perform FCI simulations of the system in the next section. These numerical simulations also extend to squeezed quantum dots, similar to that shown for a single hole in Fig.~\ref{fig:single}d-f. Even for such squeezed dots, we find a good qualitative match with this simple, effective model.

The ground-states are given by block-diagonalizing the lowest two levels of Hamiltonian~\eqref{eq:FCImatrix}. Up to energy shifts, the ground-state Hamiltonian is well-approximated by
\begin{align}
    H_\text{eff} \approx \frac{1}{2}\left[\boldsymbol{\sigma}\cdot g_{E_0}\mathbf{B} +\boldsymbol{\sigma}\cdot g_{E_1}\mathbf{B}  + \cos(\zeta)\left(\boldsymbol{\sigma}\cdot g_{E_0}\mathbf{B} -\boldsymbol{\sigma}\cdot g_{E_1}\mathbf{B}\right)\right]\,,
\end{align}
with $\zeta=\arctan(\mathcal{E}_{\text{orb},E_1}-\mathcal{E}_{\text{orb},E_0},2t_{E_0,E_1})$. The intra-dot Coulomb interaction hybridizes the spin and orbital degrees similarly to multi-dot spin-charge qubits such as the flopping-mode qubit. Consequently, we expect that the resulting system could behave similarly and that the many-body interaction may explain the enhancement of Rabi frequencies observed in the experiment.

\begin{figure}[t]
\centering
\includegraphics[width=0.5\columnwidth]{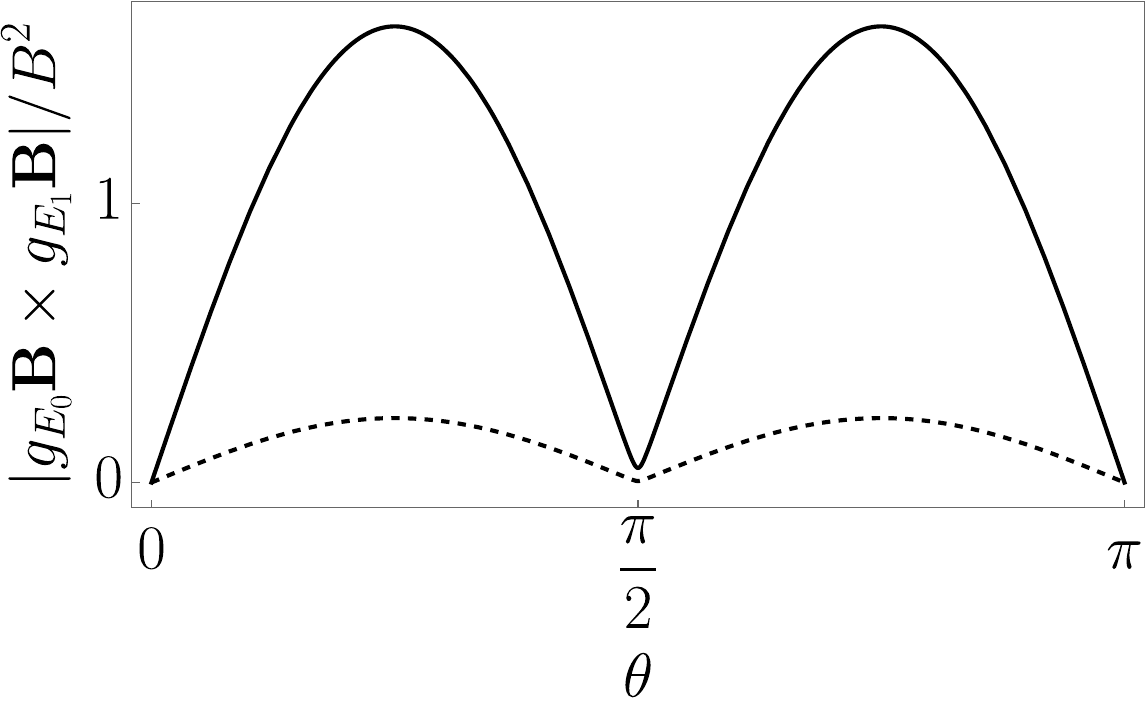}
\caption{Dependence of the driving prefactor $|g_{E_0}\mathbf{B}\times g_{E_1}\mathbf{B}|/B^2$ on the magnetic field angle $\theta$ for a (solid) isotropic and (dashed) squeezed dot and $\phi=90^\circ$. Two holes fill the ground-state $s$-orbital; in the isotropic configuration, the remaining hole of $\ket{E_0}$ and $\ket{E_1}$ occupies the $p$-orbitals $\ket{p_{[100]}}$ and $\ket{p_{[010]}}$ of a 2D harmonic oscillator, while in the squeezed configuration, the remaining hole occupies the $p$- and $d$-orbitals $\ket{p_{[100]}}$ and $\ket{p_{[200]}}$. For this calculation, we used approximate expressions for the $g$-tensors taken from Ref.~\cite{Abadillo-Uriel2023Hole-SpinInteractions}, neglecting strains and assuming separability ($\braket{p_{\xi}p_{\chi}}=\delta_{\xi,\chi}$), and the relation $\braket{p^2_{\xi}}_n=(2n+1)\braket{p^2_{\xi}}_0$, where $n$ is the $n$-th excited harmonic oscillator state in $\xi,\chi=[100],[010]$ directions. Furthermore, we used $\braket{p_{[100],[010]}^2}_0=\frac{\hbar}{2}(\unit[30]{nm})^{-2}$ for the isotropic and $\braket{p_{[100]}^2}_0=\frac{\hbar}{2}(\unit[35]{nm})^{-2}$ and $\braket{p_{[010]}^2}_0=\frac{\hbar}{2}(\unit[25]{nm})^{-2}$ for the squeezed configuration. }
\label{fig:magneticField}
\end{figure}

The Rabi frequency from driving gate $V^{(k)}\rightarrow V^{(k)}+V^{(k)}_\text{ac}$ is given by
\begin{align}
    f_R=\frac{\mu_B V^{(k)}_\text{ac}}{2h} \frac{||(g\mathbf{B})\times (\frac{d g}{d V}\mathbf{B})||}{||g\mathbf{B}||}.
\end{align}
Here the total g-tensor is given by
\begin{align}
    g= \frac{1}{2}\left[ g_{E_0} + g_{E_1}  + \cos(\zeta)\left( g_{E_0} -g_{E_1}\right)\right]
\end{align}
and the derivative with respect to gate voltages are
\begin{align}
    \frac{d g}{d V^{(k)}} =& \,\, \cos^2\left(\frac{\zeta}{2}\right) \frac{d g_{E_0}}{d V^{(k)}} + \sin^2\left(\frac{\zeta}{2}\right) \frac{d g_{E_1}}{d V^{(k)}} + \frac{1}{2}\left( g_{E_0} -g_{E_1}\right) \frac{d \cos(\zeta)}{d V^{(k)}},\\
    \frac{d \cos(\zeta)}{d V^{(k)}} =& 2\,\frac{t_{E_0,E_1}\left(\frac{d }{dV^{(k)}} \mathcal{E}_{\text{orb},E_1}-\frac{d }{dV^{(k)}} \mathcal{E}_{\text{orb},E_0}\right)+ \left(\mathcal{E}_{\text{orb},E_1}-\mathcal{E}_{\text{orb},E_0}\right)\frac{d }{dV^{(k)}}t_{E_0,E_1}}{(\mathcal{E}_{\text{orb},E_1}-\mathcal{E}_{\text{orb},E_0})^2+t_{E_0,E_1}^2}\,.
\end{align}
The Rabi frequency thus has two important contributions: A conventional g-tensor contribution arising from the sum and differences of the two individual g-tensors and a novel many-body contribution that provides a similar enhancement in driving efficiency as the flopping mode qubit. The many-body contribution can be explicitly expressed by
\begin{align}
    \mathbf{f}_R^{MB}\equiv& \frac{\mu_B V^{(k)}_\text{ac}}{8h ||g\mathbf{B}||}\left[ g_{E_0} + g_{E_1}  + \cos(\zeta)\left( g_{E_0} -g_{E_1}\right)\right]\mathbf{B} \times \left[  \frac{d \cos(\zeta)}{d V^{(k)}}\left( g_{E_0} -g_{E_1}\right)\right]\mathbf{B} \\
    =& -\frac{\mu_B V^{(k)}_\text{ac}}{4h ||g\mathbf{B}||}\frac{d \cos(\zeta)}{d V^{(k)}} g_{E_0}\mathbf{B}\times g_{E_1}\mathbf{B}.
\end{align}
Since Coulomb matrix elements and orbital energies are strongly affected by deformations (breathing) and less by lateral movement, the top plunger gate can lead to a larger Rabi frequency than nearby barrier gates. 

To better illustrate the additional effects emerging in the three-hole quantum dots, we plot in Fig.~\ref{fig:magneticField} the prefactor $|g_{E_0}\mathbf{B}\times g_{E_1}\mathbf{B}|/B^2$ of the many-body contribution as a function of the out-of-plane magnetic field angle $\theta$ ($\phi=90^\circ$). We consider an isotropic dot as well as a dot squeezed along $[010]$, and use the equations of Ref.~\cite{Abadillo-Uriel2023Hole-SpinInteractions} for $g_{E_0}$ and $g_{E_1}$, neglecting the shear strain contribution. We observe that the many-body contribution is small (but not zero) for exactly in-plane magnetic fields, maximal at $\theta=45^\circ$, and vanishes at $\theta=0^\circ$. Furthermore, we see that the novel contribution is weaker for the squeezed configuration due to the smaller induced rotation angle between $g_{E_0}\mathbf{B}$ and $g_{E_1}\mathbf{B}$ for $\phi=90^\circ$. However, this suppression is partially recovered for a magnetic field at $\phi=45^\circ$.

This many-body contribution adds to the conventional single-particle $g$-factor modulations. We further note that this conventional contribution also differs from the single-hole case as the symmetry of the occupied orbitals is not the same. The conventional contribution also experiences an enhancement that depends on the elongation of the orbitals and on the magnetic field direction~\cite{Bosco2021SqueezedPower}. While in this experiment the total Rabi frequency is enhanced, we stress that in general these two contributions can interfere constructively or destructively, and thus we cannot exclude a priori particular scenarios where the three-hole Rabi frequencies decrease compared to non-interacting holes. This highlights the complexity of the spin dynamics in the many-particle case. We provide an illustration in the next section using configuration interaction simulations that account for all contributions.

\subsubsection{FCI simulations}

\begin{figure}[t]
\centering
\includegraphics[width=0.8\columnwidth]{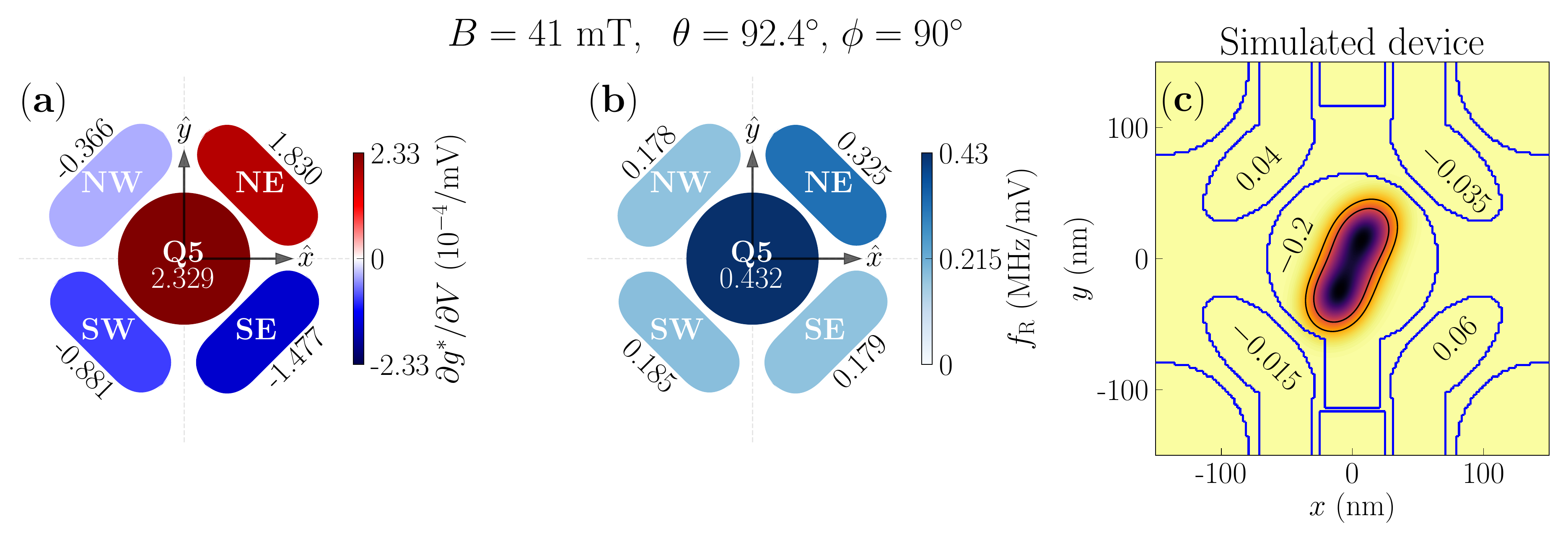}
\caption{(a) LSES, (b) Rabi frequencies $f_\mathrm{R}$ and (c) map of the density computed for three holes in a ``squeezed'' dot similar to Fig.~\ref{fig:single}d-f. The LSES of the plunger and barrier gates (per mV) are reported in panel a), while the Rabi frequencies (in MHz/mV) are reported in panel b), and the bias voltages (in V) are reported in panel c).}
\label{fig:threeholes}
\end{figure}

In our FCI simulations, we diagonalize exactly the many-body Hamiltonian [Eq.~\eqref{eq:FCImatrix}] in a basis of all Slater determinants built from the first 48 single-particle orbitals computed on the finite-differences grid (Sec.~\ref{sec:SP}) \cite{Abadillo-Uriel2021Two-bodyQubits}. We illustrate the role of Coulomb interactions on a squeezed three-hole dot similar to Fig.~\ref{fig:single}d-f. The LSES and Rabi frequencies of each gate, as well as the map of the three-hole density are plotted in Fig.~\ref{fig:threeholes}. The density is visibly more extended along the major axis of the dot than in Fig.~\ref{fig:single}f, partly because the Coulomb interactions tend to split the holes apart \cite{Abadillo-Uriel2021Two-bodyQubits}. However, the main qualitative features are the same as in the single hole case: the LSES of the plunger gate remains positive, while the LSES of the barrier gates still show a mixed blue/red pattern. Noticeably, the Rabi frequencies achieved with the plunger gate are now larger than those achieved with the barrier gates. This feature is quite generic (though not systematic); it is already prominent in the non-interacting limit, where the ground-state orbital is filled with two holes, leaving one unpaired spin in the first excited state with an approximate $p$-like envelope. This $p$-like envelope gives rise to faster Rabi oscillations than the more isotropic $s$-like envelope of the ground state because it breathes far more inhomogeneously in the field of the plunger gate. This behavior extends in the interacting regime, where Coulomb interactions moreover mix orbital configurations and thus induce additional $g$-factors modulations, as discussed in the previous section.

Our analytical and numerical models reproduce and explain many generic features of the experimental data, including the sign of the LSES of the plunger gate, the versatility of the color patterns of the LSES of the barrier gates, and the enhancement of the Rabi frequency of the plunger gate in three-hole dots. It provides insights into the physics at work in these devices. It shows, in particular, that the LSES and Rabi oscillations result from a combination of g-TMR mechanisms involving modulations of the principal g-factors as well as rotations of the gyro-magnetic axes. The strength of these mechanisms is dependent on the symmetry of the hole wave function, thus on the balance between barrier gate voltages and on disorder. The latter moreover pins the motion of the hole to some extent, and can thus significantly change the response to electrical perturbations, especially in the vicinity of the sweet lines of the barrier gates. 

%